\RequirePackage{fix-cm}
\documentclass[smallextended]{svjour3}       
\smartqed  
\usepackage{graphicx}
\usepackage[nolist]{acronym}
\usepackage{amsmath}
\usepackage{array}
\usepackage{booktabs}
\usepackage{circledtext}
\usepackage[inline]{enumitem}
\usepackage{hyperref}
\usepackage{multirow}
\usepackage{natbib}
\usepackage{xcolor}
\usepackage{xspace}
\usepackage{comment}
\usepackage{tabularx}
\usepackage{tabularx}
\usepackage{float}

\circledtextset{resize=real}

\journalname{Empirical Software Engineering}
\begin{document}

\title{Testing with Coding Agents: A Large-Scale Comparison of Agent-Supported and Human-Authored Pull Requests
}

\titlerunning{Testing with Coding Agents}

\author{Roberto Milanese \and Francesco Salzano \and Hovhannes Muradyan \and Jingjing Wu \and Remo Pareschi \and Fausto Fasano \and Mattia Fazzini
}


\institute{R. Milanese \textsuperscript{*}\at
              Politecnico di Torino, Turin, Italy \\ University of Molise, Campobasso, Italy \\
              \email{roberto.milanese@polito.it}           
           \and
           F. Salzano \at
            University of Molise, Campobasso, Italy \\
              \email{francesco.salzano@unimol.it}   
           \and
           H. Muradyan \at
            University of Minnesota, Minneapolis, USA\\
              \email{murad031@umn.edu}
           \and
           J. Wu \at
           University of Minnesota, Minneapolis, USA\\
              \email{wu000295@umn.edu}
           \and
           R. Pareschi \at
            University of Molise, Campobasso, Italy\\
              \email{remo.pareschi@unimol.it}
           \and
           F. Fasano \at
            University of Molise, Campobasso, Italy\\
              \email{fausto.fasano@unimol.it}
           \and
           M. Fazzini \at
           University of Minnesota, Minneapolis, USA\\
              \email{mfazzini@umn.edu}
}

\date{Received: date / Accepted: date}

\begin{acronym}
    \acro{AME}{average marginal effect}
    \acro{CI}{confidence interval}
    \acro{GLMM}{generalized linear mixed model}
    \acro{PR}{pull request}
\end{acronym}

\maketitle

\begin{abstract}
\textbf{Context.} Coding agents increasingly contribute tests through pull requests that developers then maintain, so how testing work is divided affects long-term maintainability.

\textbf{Objective.} We compare pull requests in which developers and coding agents collaborate (agent-supported) with pull requests without detectable agent involvement (human-authored) along five dimensions: the frequency and extent of test changes, their type and context, test structure, test smells, and code coverage.

\textbf{Method.} We analyze 114,134 pull requests merged between January 2025 and April 2026 in 391 Java repositories, identifying five coding agents through pull-request-level and commit-level signals. Since agent adoption varies across projects, we compare the two groups within repositories with mixed-effects models that control for change size.

\textbf{Results.} Pooled across repositories, both groups change test code almost equally often; within repositories, the odds that an agent-supported pull request does so are 1.65 times higher. Agent-supported pull requests add tests more often (odds ratios 1.50-1.58) and modify (0.70-0.76) and delete (0.51-0.77) them less often. We detect no difference in most structural characteristics of the tests they add, which more often have an oracle, nor, in the 95 repositories we could build, in the coverage of their production changes; in test-only pull requests, however, their tests cover less of the code they exercise. Differences in test smells are borderline and not reproduced for individual test methods.

\textbf{Conclusions.} Agent participation is associated with a different mix of testing work, which pooled comparisons across projects can hide or reverse; its relationship with test quality remains uncertain.
\end{abstract}

\section{Introduction}\label{sec1}

AI-based coding tools have evolved from completing individual lines of code to carrying out entire development tasks: modern coding agents iterate over a task, invoking build systems, test frameworks, and version control until they consider it complete~\citep{robbes2026b, wang2025ai}. In collaborative projects, the result usually reaches the code base as a \ac{PR} that developers review, revise, and integrate, and adoption has been rapid: as of February 2026, between 22.20\% and 28.66\% of the 128,018 active GitHub repositories analyzed by \citet{robbes2026b} showed traces of coding agent use. We refer to \acp{PR} in which developers and coding agents both participate as \emph{agent-supported} \acp{PR}, and to \acp{PR} without detectable agent involvement as \emph{human-authored} \acp{PR}.

Testing is where the consequences of this shift are likely to last longest. Tests are read, modified, and deleted as the code they exercise evolves, and design problems in test code, known as test smells, make them harder to understand and change~\citep{bavota2012empirical}; such problems are usually introduced when a test is first written and rarely removed afterwards~\citep{tufano2016empirical}. As agents write tests that developers will then maintain, two questions become central: what kind of testing work agents take on, and what kind of test code results.

Research on agentic testing has grown quickly, but the evidence is fragmented. Studies of tests in real agentic \acp{PR} each focus on one aspect: test inclusion and evolution~\citep{haque2026empirical}, test structure and coverage~\citep{yoshimoto2026testing}, diff coverage and assertion quality~\citep{alves2026test}, or test failures~\citep{hoseinpour2026empirical}. Only some compare agentic with human contributions, and they pool them across repositories, at least in part, or analyze few projects one by one (Section~\ref{sec3}). What is missing is a comparison over many repositories that estimates the difference within each of them, controls for the size of the contribution, and quantifies how much the difference varies between repositories. This matters because agent adoption is uneven across projects: a pooled comparison mixes whether agentic contributions test differently with which projects use agents, and, as we show, the second can hide and even reverse the first.

Our previous study~\citep{milanese2026} compared 6,582 agent-supported and 3,122 human-authored \acp{PR} from the \textsc{AIDev} dataset~\citep{li2025a} in four programming languages. It found that the two groups include tests at comparable rates (42.9\% against 40.0\%), but that agent-supported \acp{PR} devote a larger share of their changes to test code; that during co-evolution they more often add tests (OR 1.79), whereas human-authored \acp{PR} more often modify and delete existing ones; and that differences in test smells were negligible in magnitude. Its design, however, compared \acp{PR} pooled across repositories, classified changes at the level of test files, considered any \ac{PR} that changes both test and production code as co-evolution, and did not measure coverage.

This paper addresses those limitations. First, we compare the two groups within repositories, using mixed-effects models in which each repository has its own baseline and its own difference between the groups, controlling for the size of the change. Second, we analyze test changes at the level of test methods and fixtures, and consider a test change as co-evolving only when it is linked to production code changed by the same \ac{PR}~\citep{marsavina2014}; we also detect test smells with \textsc{JNose}~\citep{virginio2020jnose}, which locates each occurrence in a test method, so that smells can be compared both for entire contributions and for individual tests. Third, we add two dimensions: the structure of test code, extracted with \textsc{Hamster}~\citep{pan2026}, and code coverage, measured with \textsc{JaCoCo} by executing each test suite before and after the \ac{PR}. Since measuring coverage requires building each project, we restrict the study to Java, whose standardized build tools make this feasible at scale (Section~\ref{sec:study-context}).

These choices required a new dataset. We collected all \acp{PR} created and merged between January 1, 2025, and April 30, 2026, in Java repositories with traces of agent activity, identifying agent participation through \ac{PR}-level and commit-level signals~\citep{robbes2026c} for five widely adopted agents: Claude Code, Codex, Copilot, Cursor, and Devin. Retaining the repositories with at least ten \acp{PR} of each group, we obtain 18,375 agent-supported and 95,759 human-authored \acp{PR} from 391 repositories. On this dataset, we address five research questions (Section~\ref{sec4}), which compare the two groups in the frequency and extent of testing (RQ$_1$), the type and context of test changes (RQ$_2$), the structure of test code (RQ$_3$), test smells (RQ$_4$), and code coverage (RQ$_5$).

Our results lead to five main findings. First, how the comparison is made determines what it shows: pooled across repositories, the two groups change test code almost equally often (60.1\% against 59.4\%), whereas within the same repository and at equal size, the odds that an agent-supported \ac{PR} changes test code are 1.65 times those of a human-authored one. Second, relative to human-authored \acp{PR} of the same size, agent-supported \acp{PR} add tests more often (OR 1.50-1.58) and modify (OR 0.70-0.76) and delete (OR 0.51-0.77) them less often, in every development context, while the two groups differ little in how often they co-evolve tests with production code (OR 1.12). Third, we detect no difference in most structural characteristics of the tests that agent-supported \acp{PR} add, except that they more often have an oracle (OR 1.71); when modifying tests, agent-supported \acp{PR} more often add assertions (OR 1.31). Fourth, the evidence on test smells is weaker: at the \ac{PR} level, agent-supported \acp{PR} are associated with a smaller net increase in smells when adding tests and a larger one when modifying them, but both estimates are borderline and neither is reproduced for individual test methods. Fifth, we detect no difference in how well the two groups cover their production changes, but in test-only \acp{PR} the tests that agent-supported \acp{PR} contribute cover less of the classes they exercise ($-$11.1 percentage points), a result that rests on only 11 repositories with test-only \acp{PR} of both groups. For most outcomes, the size and sometimes the direction of the difference vary considerably between repositories.

In summary, this paper makes the following contributions:

\begin{itemize}
    \item a dataset of 114,134 merged \acp{PR} from 391 Java repositories, each containing both agent-supported and human-authored \acp{PR}, in which agent participation is identified through multiple \ac{PR}-level and commit-level signals for five coding agents;
    \item a within-repository comparison of testing in the two groups across 391 repositories, which controls for change size, estimates how much the difference varies between repositories, and shows that pooling contributions across projects can conceal or reverse it;
    \item an analysis of the same contributions along five testing dimensions, including test structure and coverage, with test changes classified at the level of test methods and linked to the production changes they accompany, and test smells compared at the level of contributions and of individual test methods;
    \item a replication package containing the dataset, the queries, and the analysis scripts~\citep{milanese2026r}.
\end{itemize}

The remainder of this paper is organized as follows. Section~\ref{sec2} introduces background on coding agents, testing in pull-request-based development, and test smells, and Section~\ref{sec3} discusses related work. Section~\ref{sec4} describes the study design. Section~\ref{sec:results} reports the results, which Section~\ref{sec:discussion} discusses together with the threats to validity. Section~\ref{sec7} concludes the paper and outlines future work.

\section{Background}\label{sec2}
\subsection{AI Coding Agents and Agent-Supported Pull Requests}

Unlike code completion assistants, which help developers write individual lines or blocks of code, \emph{coding agents} undertake entire development tasks, such as fixing bugs or implementing features~\citep{robbes2026b}. A coding agent can be characterized as an LLM operating in an iterative loop with access to external tools: its actions are executed, their outcomes are fed back to the model, and the process continues until the task is considered complete. Agentic programming is therefore goal-oriented, multi-step, and tool-based: agents interact with compilers, debuggers, test frameworks, and version control systems, while developers may retain oversight by providing goals, constraints, or feedback~\citep{wang2025ai}.

In collaborative development workflows, this interaction materializes through pull requests, where coding agents may produce commits or entire contributions that are subsequently inspected, refined, and integrated by human developers~\citep{robbes2026b}. The division of labor within such a contribution is not recorded explicitly, but agent participation usually leaves observable traces, such as dedicated branch names, bot accounts, or co-authorship lines in commit metadata, which is what makes it possible to study these contributions at scale.

We distinguish \emph{agent-supported} \acp{PR}, in which developers and one or more coding agents participate, from \emph{human-authored} \acp{PR}, produced without the involvement of coding agents that leaves detectable traces. Our previous study called them human-agent collaborative pull requests and human pull requests~\citep{milanese2026}. Since agent participation is inferred from its traces, the second group is primarily rather than exclusively human-authored (Section~\ref{sec:study-context}).

\subsection{Testing in Pull-Request-Based Development}

In real-world development, tests are a persistent part of a project's quality-assurance infrastructure and evolve together with the code base, so they need to remain understandable, maintainable, and aligned with the software they exercise~\citep{pan2026}. This has become more relevant as coding agents contribute tests that are then maintained within long-lived test suites.

Within pull-request-based development, testing takes different forms depending on the scope of the contribution. Some \acp{PR} modify only production code, others change production and test code together, a setting commonly described as \emph{test-production co-evolution}, and others focus exclusively on test code.

Within a contribution, test code evolves through a small set of recurring tasks. Following~\citet{pinto2012}, a contribution can \emph{add} new tests, \emph{modify} existing ones, or \emph{delete} them, and these tasks place different demands on the contributor: adding a test is a largely self-contained activity, whereas modifying or deleting one requires understanding what an existing test checks and why it should change. The structural notion of co-evolution introduced above is also coarse, since a pull request may well change test and production code that have nothing to do with each other. A finer definition considers a test change to co-evolve only when it is linked to production code changed by the same contribution, for example because the test exercises a method that the contribution modifies~\citep{marsavina2014}. Combining tasks with the context in which they occur therefore characterizes testing work more precisely than the presence of test changes alone.

These dimensions describe how testing work is distributed within a contribution, but not the quality of its outcome. Two further properties are commonly used for that purpose. The first is the internal quality of the test code itself, which we discuss in the next subsection. The second is test adequacy, usually approximated by code coverage, that is, the share of production code that the test suite exercises. The two are complementary: a contribution may raise coverage while leaving test code that is hard to maintain, and well-designed tests may exercise little of the code they accompany.

\subsection{Test Smells and Test-Code Quality}

Test code, like production code, may exhibit design and implementation choices that hinder its maintenance. Such recurring choices are known as \emph{test smells}: patterns associated with poorly designed tests and with the way test cases are organized, implemented, and interact with one another~\citep{bavota2012empirical}. Typical examples include a test that verifies several unrelated behaviors, a group of assertions that gives no indication of which one failed, or a fixture shared by a class but used by only some of its tests. A smelly test is not necessarily incorrect: it may detect faults reliably and still be hard to read, to change, or to keep independent of conditions unrelated to the behavior it checks, which is why test smells are a quality concern distinct from test effectiveness.

These costs accumulate because test suites are long-lived. \citet{bavota2012empirical} found test smells to be widespread in Java systems and showed experimentally that several categories hinder comprehension, while~\citet{tufano2016empirical} observed that smells are usually introduced when a test class is first committed and rarely removed afterwards. The quality of a test is therefore largely decided by the contribution that writes it, and the cost of a poor design is paid by every later contribution that touches it.

Test smells are detected statically, by matching structural rules against the source code of a test class, so detectors report design properties rather than defects, and the resulting counts approximate test-code quality rather than measure it. Since they can be computed on any revision, they allow comparing the test code before and after a contribution and attributing the difference to that contribution.

\section{Related Work}\label{sec3}
A broad body of research has investigated coding assistants and agents across different software engineering activities. \cite{nikolaidis2024comparison}, for example, compared \textsc{ChatGPT} and \textsc{Copilot} for Python code generation, focusing on the correctness and quality of the produced code. \cite{sergeyuk2025using} studied the practical adoption of coding agents, examining developer perceptions, usage patterns, and organizational implications. Similarly, \cite{butler2025dear} evaluated generative AI coding tools in a controlled workplace study, reporting their effects on developer productivity and behavior. However, these studies consider testing only marginally, if at all, and do not investigate how testing activities emerge, evolve, and are distributed within collaborative development workflows involving both humans and coding agents.

Closer to our setting, several studies characterized agentic \acp{PR} on the \textsc{AIDev} dataset \citep{li2025a, li2026a}. Agentic \acp{PR} are accepted less often than human ones \citep{li2025a}, and merged agentic \acp{PR} differ from merged human \acp{PR} in the number of commits, files touched, and lines deleted \citep{ogenrwot2026modify}, while unmerged ones are larger and more often fail CI \citep{ehsani2026fail}. These size differences motivate controlling for change size. \cite{sawada2026maintenance} found that agent-generated files receive less frequent maintenance than human-authored code and that human developers perform most of it, which is why the tests that developers inherit from agents deserve attention. At the repository level, \citet{he2026cursor} compared \textsc{Cursor}-adopting projects with matched controls in a difference-in-differences design and found a large but transient gain in velocity together with a persistent increase in static-analysis warnings and complexity, a within-project comparison in the same spirit as the present study. Still, none of these studies examined the tests a \ac{PR} contributes.

Testing practices within real-world agentic development workflows have been examined from different angles. One line of work focuses on how tests appear and evolve during pull-request development. Using the AIDev dataset, \citet{haque2026empirical} studied test inclusion, the stage at which tests are introduced, and the differences between test-containing and non-test contributions. Their results indicate that testing has become increasingly common over time and that pull requests involving tests tend to be larger and require longer turnaround times. In addition, tests introduced in the initial contribution are often revised in later commits, highlighting that testing activities may evolve throughout the pull-request lifecycle.

A different perspective is provided by \citet{yoshimoto2026testing}, who investigated the prevalence of AI-authored test additions, the structural characteristics of generated tests, and their impact on code coverage. AI-generated tests were found to be slightly longer, to contain more assertions, and to rely on simpler control-flow structures than human-written tests, while achieving comparable coverage improvements. Coverage and assertion quality were further examined by \citet{alves2026test} through a comparison of AI-only, human-AI coauthored, and human pull requests. Their analysis shows that AI-only contributions achieve higher diff-coverage of modified production code, whereas assertion quality remains broadly comparable across the three authorship settings. Coverage was also the focus of \citet{dipongkor2026coverage}, who studied 4,882 agentic \acp{PR} in Java and Python and measured, for each merged \ac{PR}, the share of added executable lines covered by the existing tests and by the tests the \ac{PR} itself changes. Existing tests covered 61.5\% of these lines in Java and 27.0\% in Python, and agent-written tests raised diff coverage in only 35.9\% of the Java and 22.5\% of the Python \acp{PR} that change both code and tests.

Other work has focused on failures rather than coverage or test structure. \citet{hoseinpour2026empirical} manually analyzed 500 pull requests affected by test-related failures and reported that runtime failures occur more frequently than compile-time ones, with assertion failures and unexpected exceptions as the most common categories.

The agentic testing studies differ along the aspects that matter most for the present work: the population analyzed, the unit of analysis, the testing dimensions considered, whether agentic contributions are compared with human ones, and how the repository is taken into account. Three of them analyze agentic contributions only. Among those that compare agentic and human contributions, \citet{yoshimoto2026testing} study a small number of projects, pooling them for some analyses and reporting others per project, \citet{alves2026test} sample the human \acp{PR} from the same repositories as the agentic ones but compare the groups with pooled tests, and our previous study compared \acp{PR} pooled across repositories~\citep{milanese2026}. None of them estimates the difference within each of many repositories while controlling for the size of the contribution, or estimates how much the difference varies between repositories, which is the comparison this paper adds.

For the individual dimensions we compare, we also build on studies of testing outside agentic \acp{PR}. How developers maintain test suites is documented in studies that predate coding agents. \citet{pinto2012} found, across many versions of six Java programs, that most test changes are additions, and that tests that fail after a change are more often deleted than repaired, usually because the code they exercised no longer exists. \citet{levin2017coevolution} found, across 61 open-source Java projects, that in half of them fewer than a quarter of the bug-fixing commits also touched tests. Whether coding agents follow the same pattern is so far known only from a benchmark: \citet{shang2026breaking} found that Claude Code, Codex CLI, and OpenCode repair tests that fail but perform worst on tests that still pass yet no longer check the updated behavior, because their loop is driven by execution failures. How agent-supported contributions maintain tests in real projects, compared with human-authored ones, remains open.

The tests that AI coding assistants generate have mostly been assessed for their quality, often against developer-written tests. \cite{alves2024detecting} examined test smells in Python tests produced by \textsc{Copilot}, with a particular focus on the maintainability of the generated test code. At a larger scale, \citet{ouedraogo2026smells} compared class-level test suites generated by four LLMs with hundreds of thousands of developer-written tests from open-source Java projects, using \textsc{TsDetect} and \textsc{JNose}, and found that LLM-generated tests consistently exhibit smells such as Assertion Roulette and Magic Number Test. \cite{joshi2025disrupting} instead evaluated \textsc{Copilot}-generated tests for existing software projects by comparing them with manually developed test suites. Although these works provide useful evidence on the quality and characteristics of automatically generated tests, they largely treat test generation as an isolated activity performed on already implemented software, rather than as an integral part of the development process.

\citet{pan2026} presented \textit{Hamster}, a characterization of 1.7 million developer-written Java test cases from 1,908 open-source projects, covering fixtures, structured inputs, test scope, call-assertion sequences, mocking, and code coverage. Developer-written tests often exercise several focal classes and methods, rely on reusable fixtures and mocks, and use richer assertion patterns, whereas tests generated by EvoSuite~\citep{fraser2011evosuite} and by the LLM-based \textit{aster}~\citep{pan2025aster} are simpler and more narrowly scoped, which suggests that coverage alone does not capture how closely generated tests reflect development practice.

How the coverage of a change, rather than of a whole program, should be measured has its own literature. \citet{hilton2018coverage} analyzed 7,816 builds of 47 projects and found that, in large and mature projects, the change in overall statement coverage does not capture the nuances of code evolution; they proposed measures that separate the coverage of the patch from that of the non-patch, and changes in coverage from changes in the set of statements covered. A follow-up on nine mature projects over a decade reported that overall coverage has risen while many individual patches remain poorly tested~\citep{bailey2025evolution}. The agentic coverage studies above measured the diff coverage of the lines a \ac{PR} changes, by the tests the \ac{PR} modifies~\citep{alves2026test} or by existing and agent-written tests separately~\citep{dipongkor2026coverage}. RQ$_5$ makes the same distinction between the patch and the rest of the code, and extends line-level diff coverage with the coverage of the classes a \ac{PR} affects, before and after the change, for agent-supported and human-authored \acp{PR} from the same repositories (Section~\ref{sec:rq-5-experimental-procedure}).

\section{Study Design}\label{sec4}

The \emph{goal} of the study is to compare the testing work carried out in agent-supported and human-authored \acp{PR} of the same repositories, with the \emph{purpose} of understanding what kind of testing work coding agents take on and what test code results. We address the following research questions:

\begin{itemize}
    \item \textbf{RQ$_1$:} Are there differences between agent-supported and human-authored \acp{PR} in the frequency and extent of testing contributions?
    \item \textbf{RQ$_2$:} How do the development contexts in which testing contributions appear compare between agent-supported and human-authored \acp{PR}?
    \item \textbf{RQ$_3$:} How do the structural characteristics of the test code produced by agent-supported and human-authored \acp{PR} compare?
    \item \textbf{RQ$_4$:} How does test quality, as reflected by test smells, differ between agent-supported and human-authored \acp{PR}?
    \item \textbf{RQ$_5$:} How do agent-supported and human-authored \acp{PR} compare in terms of code coverage changes and adequacy?
\end{itemize}

Section~\ref{sec:study-context} explains why we restrict the study to Java and build a new dataset, Section~\ref{sec:data-collection} describes how we collected it, and Section~\ref{sec:experimental-procedure} the procedure we follow for each research question.

\subsection{Study Context}
\label{sec:study-context}

This work extends our previous study~\citep{milanese2026}, which characterized testing in agent-supported and human-authored \acp{PR} from the \textsc{AIDev} dataset~\citep{li2026a} in four programming languages in terms of frequency, extent, development context, and test smells. The two new dimensions, code coverage and the structure of test code, required two changes to the design: restricting the scope to Java repositories and creating a new dataset of \acp{PR}.

The decision to focus on Java projects stems from the nature of the new analyses. While test smell detection operates on static file contents, measuring coverage requires building each project and executing its test suite, and this is only feasible at scale when build and test workflows are standardized. In the Java ecosystem, \textsc{Maven}\footnote{\url{https://maven.apache.org/}} and \textsc{Gradle}\footnote{\url{https://gradle.org/}} are the build systems used by the majority of projects~\citep{jetbrains2025}, and provide a single entry point to run tests, regardless of the underlying test framework used. \textsc{JaCoCo}\footnote{\url{https://jacoco.org/jacoco/}} easily integrates with both these tools to produce widely used, well-defined coverage metrics. This contrasts with the Python and JavaScript/TypeScript ecosystems, where testing is fragmented across various frameworks~\citep{jetbrains2024a, jetbrains2024b}. Extending the study to all four languages would require separate toolchains, each with different build systems, coverage semantics, and characterization tooling, which would undermine the comparability of results.

While \textsc{AIDev} is designed to support broad studies of agent adoption across languages and ecosystems, our design requires a large sample of Java projects containing both agent-supported and human-authored \acp{PR}. In our previous work \citep{milanese2026}, we used \textsc{AIDev v3}~\citep{li2025a}, which contains 3,481 Java repositories. However, detailed \ac{PR}-level data is only available for the \textsc{AIDev-pop} subset, which includes only 86 Java repositories with more than 100 stars. Furthermore, human-authored \acp{PR} are limited to repositories with more than 500 stars, leaving only 50 qualifying Java repositories. In our previous study, we applied this same threshold to agent-supported \ac{PR} as well, to ensure that both groups came from the same sample. The more recent version of the dataset, \textsc{AIDev v4}~\citep{li2026b}, covers a larger number of Java projects (9,719 in total, 238 of which have more than 100 stars), but it does not provide any human-authored \acp{PR}, precluding any comparison. In addition to limiting the sample size, star-based thresholds are an imperfect proxy for representativeness because they capture popularity rather than the extent to which a project reflects typical development practices. For this reason, we adopt different repository selection criteria, described in Section~\ref{sec:repository-selection}.

The need to effectively identify agent-supported \acp{PR} also motivated the creation of a new dataset. \textsc{AIDev} relies on branch name prefixes for all agents except for \textsc{Claude Code}, for which it relies on the \texttt{"Co-Authored-By: Claude"} line in the \ac{PR} body~\citep{li2025b}. However, depending on a single signal per agent makes identification sensitive to project conventions. Building on the heuristics proposed by \citet{robbes2026c}, we combine \ac{PR}-level signals (branch name prefix, label, agentic author, and agentic assignee) with commit-level signals (agentic author or co-author), thereby reducing reliance on a single signal. We focus on contributions from \textsc{Claude Code}, \textsc{Codex}, \textsc{Copilot}, \textsc{Cursor}, and \textsc{Devin}, as these are the only coding agents with significant adoption in Java projects (Table~\ref{tab:java-agent-adoption}). As in our previous study, we describe the comparison group as \textit{primarily} human-authored since we cannot exclude agent use that leaves none of these signals.

\begin{table}[t]
    \centering
    \caption{Number of Java \acp{PR} per coding agent identified using branch name prefixes, following the heuristics from \citet{robbes2026c}. Retrieved from the GitHub Search API on May 10, 2026 using the query: \texttt{is:pr is:closed head:\{term\} language:java draft:false created:2025-01-01..2026-04-30 closed:2025-01-01..2026-04-30 sort:created}.}
    \begin{tabular}{llr}
        \toprule
        Coding Agent & Branch Name Prefix & \# \acp{PR} \\
        \midrule
        \textbf{Codex} & \texttt{codex/} & 98,300 \\
        \textbf{Copilot} & \texttt{copilot/} & 36,600 \\
        \textbf{Claude Code} & \texttt{claude/} & 18,800 \\
        \textbf{Devin} & \texttt{devin/} & 2,900 \\
        \textbf{Cursor} & \texttt{cursor/} & 2,700 \\
        \midrule
        Jules & \texttt{jules/} & 178 \\
        Codegen & codegen-\texttt{bot/} & 74 \\
        Opencode & \texttt{opencode/} & 43 \\
        Coderabbit & \texttt{coderabbit/} & 11 \\
        Serena & \texttt{serena/} & 9 \\
        Sweep & \texttt{sweep/} & 8 \\
        Gru & \texttt{gru/} & 5 \\
        Open SWE & open-\texttt{swe/} & 5 \\
        OpenHands & \texttt{openhands/} & 2 \\
        \bottomrule
    \end{tabular}
    \label{tab:java-agent-adoption}
\end{table}

\subsection{Data Collection}
\label{sec:data-collection}

\begin{figure}
    \centering
    \includegraphics[width=0.95\linewidth]{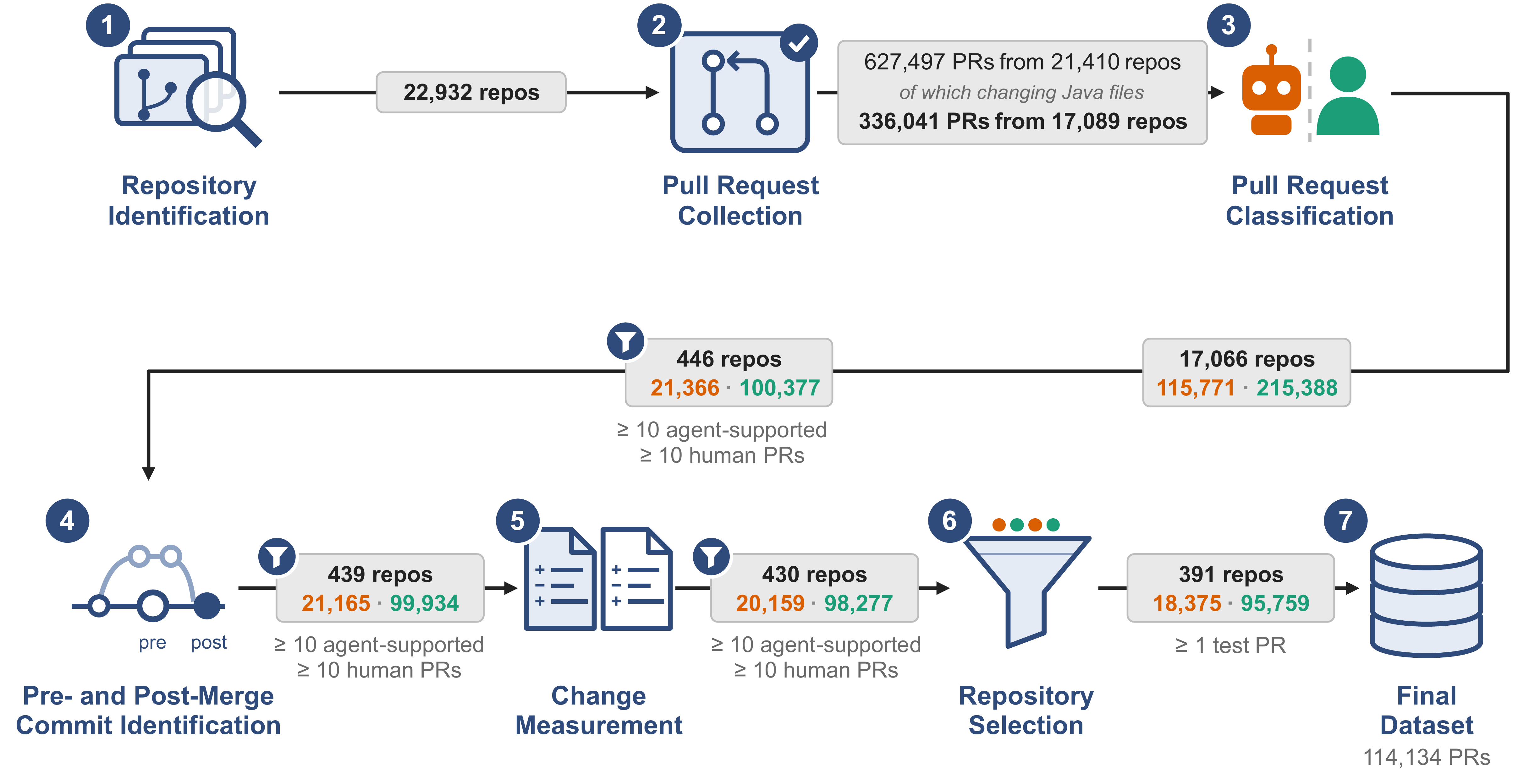}
    \caption{Summary of the data collection pipeline. Grey boxes report the number of repositories, and of agent-supported (orange) and human-authored (green) \acp{PR} after each step}
    \label{fig:data-collection}
\end{figure}

Figure~\ref{fig:data-collection} summarizes our data collection pipeline. First, we identify Java repositories with at least one agent-supported contribution \circledtext*[width=.9em,charshrink=.65]{1}, and retrieve all their \acp{PR} that were created and merged between January 1, 2025, and April 30, 2026 \circledtext*[width=.9em,charshrink=.65]{2}. Next, we classify each \ac{PR} as either agent-supported or human-authored \circledtext*[width=.9em,charshrink=.65]{3}, identify its pre- and post-merge commits \circledtext*[width=.9em,charshrink=.65]{4}, and measure the size of the changes it introduces to the test and production code \circledtext*[width=.9em,charshrink=.65]{5}. Finally, we retain repositories containing enough \acp{PR} of both types to support a fair within-repository comparison \circledtext*[width=.9em,charshrink=.65]{6}. This results in our final dataset \circledtext*[width=.9em,charshrink=.65]{7}.

\subsubsection{Repository Identification}

Since most GitHub repositories still show no traces of coding agent use~\citep{robbes2026b}, rather than covering every Java repository, we first identify those containing at least one agent-supported contribution and then retrieve all their merged \acp{PR}. To this end, we query the GitHub Search API using the signals proposed by \citet{robbes2026c} for the five coding agents within our scope, reported in Table~\ref{tab:agent-patterns}. This results in a total of 25 queries, available in our replication package \citep{milanese2026r}.

At the \ac{PR} level, we use the GraphQL API\footnote{\url{https://docs.github.com/en/graphql/}} with the \texttt{language:java} qualifier to search for non-draft \acp{PR} created and closed within our time window whose head branch name starts with an agent-specific prefix, carry an agent-specific label, or have an agent account as author or assignee. At the commit level, we search for commits authored within our time window whose author email address or login belongs to an agent; since commit searches are only available through the REST API\footnote{\url{https://docs.github.com/en/rest/}} and do not support the \texttt{language} qualifier, we retain the repositories whose primary language, as detected by \textsc{Linguist}\footnote{\url{https://github.com/github-linguist/linguist/}}, is Java.

Since the Search API returns at most 1,000 results per query, we recursively split the time range of any query that exceeds this limit, by creation date for \ac{PR} queries and author date for commit queries, until each interval is entirely retrieved, as in \cite{romano2021}.

\begin{table}[t]
    \centering
    \caption{Signals used to discover candidate repositories for each coding agent. \ac{PR}-level signals are searched among closed, non-draft \acp{PR} with the \texttt{language:java} qualifier. Commit-level signals match the email address or login of a commit's author.}
    \begin{tabular}{>{\raggedright\arraybackslash}p{1.7cm} >{\centering\arraybackslash}p{1.65cm} >{\centering\arraybackslash}p{.75cm} >{\centering\arraybackslash}p{1.95cm} >{\centering\arraybackslash}p{3.72cm}}
        \toprule
        Coding Agent & Branch Name Prefix & \ac{PR} Label & \ac{PR} Author or Assignee & Commit Author \\
        \midrule
        Claude Code & \texttt{claude/}, \texttt{claudecode/} & – & \texttt{claude} & \texttt{assistant@anthropic.com}, \texttt{claude@anthropic.com}, \texttt{noreply@anthropic.com} \\[25pt]
        Codex & \texttt{codex/} & \texttt{codex} & \texttt{codex} & \texttt{codex@openai.com} \\[5pt]
        Copilot & \texttt{copilot/} & – & \texttt{app/copilot-} \texttt{swe-agent} & \texttt{copilot-swe-agent[bot]}, \texttt{copilot@github.com} \\[15pt]
        Cursor & \texttt{cursor/} & – & \texttt{cursor} & \texttt{cursoragent@cursor.com} \\[5pt]
        Devin & \texttt{devin/} & – & \texttt{app/devin-ai-} \texttt{integration} & \texttt{devin-ai-integration[bot]} \\
        \bottomrule
    \end{tabular}
    \label{tab:agent-patterns}
\end{table}

Overall, we identify 22,932 Java repositories with at least one candidate agent-supported contribution.

\subsubsection{Pull Request Collection}

Identifying a repository with agentic activity is not enough, because commit-level matches are not linked to the \acp{PR} that contain them, and the human-authored \acp{PR} needed for our comparison are not matched at all. Since the API endpoints that list the \acp{PR} of a repository do not support filtering by date, we use the Search API again to retrieve all \acp{PR} created and closed within our time window in each candidate repository, together with their head and target branches, merge commit, labels, author, commits with their authors and co-authors, and changed files. The complete query is available in our replication package \citep{milanese2026r}.

We retain only merged \acp{PR}, for two reasons: they are the contributions that become part of the code base after review, and thus reflect the testing practices that projects accept, and most of our analyses compare the repository before and after a \ac{PR} is integrated, which is only possible for merged \acp{PR}.

In total, we collect 627,497 merged \acp{PR} from 21,410 repositories. As an initial filter, we discard \acp{PR} with no Java files in the changed files list, leaving 336,041 \acp{PR} from 17,089 repositories. We further refine this filter in Section~\ref{sec:change-measurement} by measuring the actual Java code change size in each \ac{PR}, i.e., the number of added, modified, and deleted lines of Java code, excluding comments and blank lines.

\subsubsection{Pull Request Classification}

After collecting \acp{PR}, we classify them as either agent-supported or primarily human-authored using the most recent version of the heuristics released by \cite{robbes2026a}. These signals cover more than 40 coding agents, including the five within our scope, with agent-specific branch name prefixes, labels, and author identities (i.e., account logins, names, and email addresses).

Classification proceeds in two tiers. First, at the \ac{PR} level, we check if the head branch name, one of the labels, or the author matches an agent pattern. We do not consider assignees at this stage, since an agent can be assigned to a \ac{PR} without contributing code, such as when a developer asks \textsc{Copilot} to review their changes. If no \ac{PR}-level signal applies, we check if the author or a co-author of any of its commits matches an agent pattern. A commit-level match only counts if the matching commit modifies at least one Java file. Therefore, an agent that only edited documentation or build files would not make a \ac{PR} agent-supported for the purposes of our study.

\acp{PR} that do not match any of the five agents in our scope, but that do match other agents covered in \cite{robbes2026a} are discarded because they fall outside our scope. If no agent is matched at all, we check if any commit that modifies a Java file was authored by a bot account (i.e., an account whose login or name contains \texttt{[bot]} or \texttt{-bot}). These are discarded as well, as their automated changes would distort the human-authored group, and all remaining \acp{PR} are considered human-authored.

Out of the 336,041 merged Java \acp{PR} collected in the previous step, 115,771 are agent-supported, 215,388 are human-authored, 2,386 are attributed to other agents, and 2,496 are attributed to bots. Most agent-supported \acp{PR} (115,398) match a single agent, while 373 match two or more.

\subsubsection{Pre- and Post-Merge Commit Identification}
\label{sec:pre-post-merge-commit}

In order to accurately measure the size of the changes introduced by a \ac{PR}, we need to extract two commits: the pre-merge commit (i.e., the latest commit on the target branch before the \ac{PR} is merged) and the post-merge commit (i.e., the latest commit on the target branch after the \ac{PR} is merged).

\begin{figure}
    \centering
    \includegraphics[width=.85\linewidth]{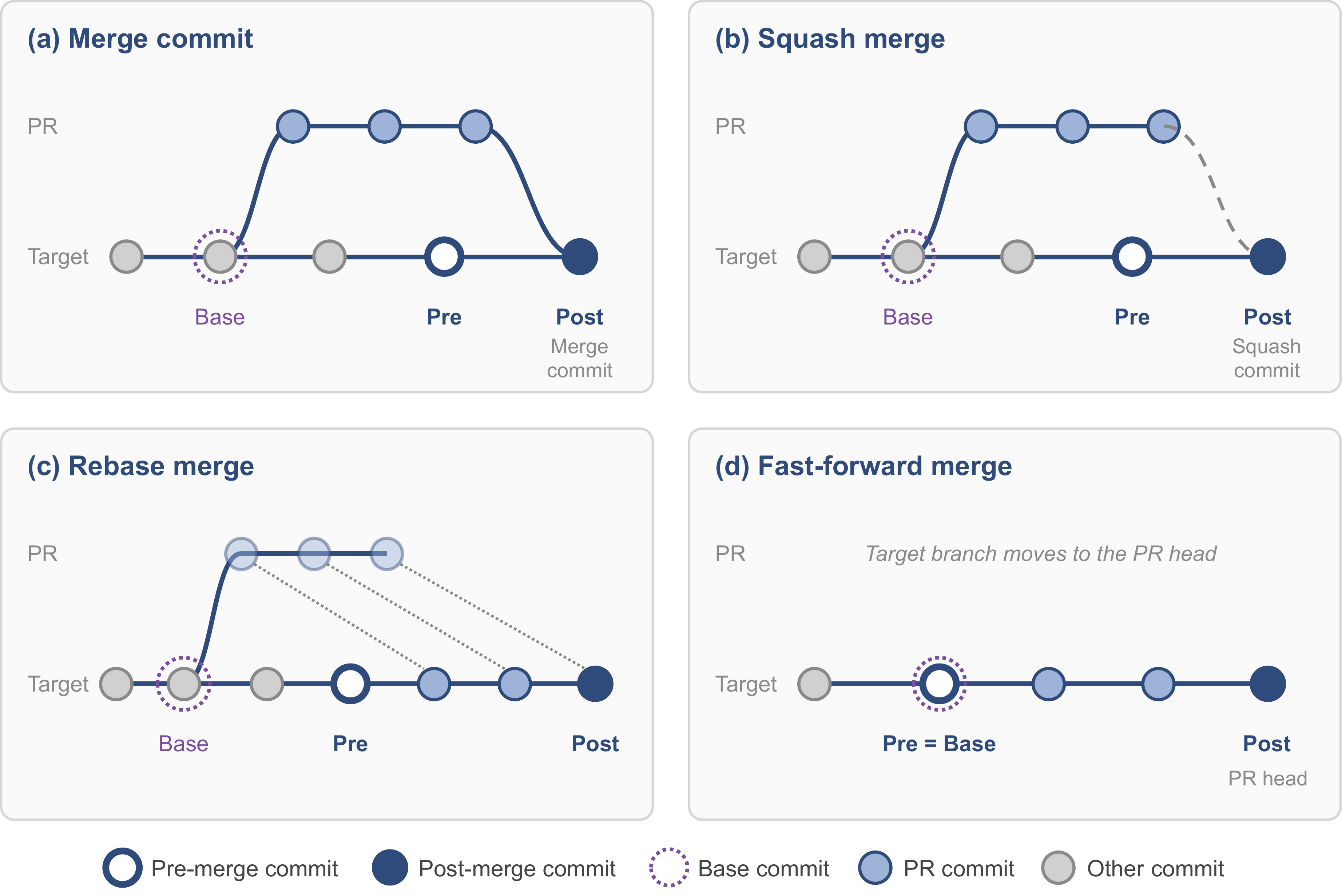}
    \caption{Pre- and post-merge commits under each merge strategy. The pre-merge commit is the most recent commit on the target branch before the \ac{PR} is merged. GitHub’s base commit may precede it if the target branch advances in the meantime, as in (a)-(c)}
    \label{fig:pre-post-merge-commit}
\end{figure}

We cannot rely on the base commit reported by the GitHub API, because it reflects the state of the target branch when the \ac{PR} was opened, and comparing it with the post-merge commit would attribute to the \ac{PR} any change merged into the target branch in the meantime. GitHub instead always records the post-merge commit as the merge commit of the \ac{PR}, which, depending on the merge strategy, is the merge commit itself, the squash commit, the last rebased commit, or the head of the \ac{PR} for fast-forward merges. Comparing the two commits reveals the net change that the \ac{PR} introduced, including changes made while resolving merge conflicts (Figure~\ref{fig:pre-post-merge-commit}). We therefore clone each repository, fetch any missing references to \ac{PR} heads, including those opened from forks, and derive the pre-merge commit from the merge strategy: it is the first parent of the post-merge commit for merge commits and squash merges (Figure~\hyperref[fig:pre-post-merge-commit]{2a-b}); the \textit{N}-th ancestor of the post-merge commit for rebase merges, which rewrite the \textit{N} commits of a \ac{PR} on top of the target branch and whose rewritten commits we match to the original ones by author identity, author date, and message (Figure~\hyperref[fig:pre-post-merge-commit]{2c}); and the first commit in the history that does not belong to the \ac{PR} for fast-forward merges (Figure~\hyperref[fig:pre-post-merge-commit]{2d}).
We exclude \acp{PR} for which the pre- and post-merge commits cannot be identified (e.g., when the merge commit is no longer tracked by GitHub). Some \acp{PR} may also share the same merge commit if they are opened from the same branch. If they are all either agent-supported or human-authored, we keep the one that was actually merged from GitHub, otherwise we exclude them all, since the change cannot be accurately attributed to either group.

\subsubsection{Change Measurement}
\label{sec:change-measurement}

After identifying the pre- and post-merge commits, we compare them to measure the size of the changes that the \ac{PR} introduced to the Java code. This is related to \textit{code churn}, which is typically calculated as the sum of added and deleted lines of code \citep{nagappan2005, gousios2014}. To better reflect the actual size of a change rather than the number of line edits, we instead count each modified line once, rather than counting it as both a deletion and an addition.

First, we use the \texttt{git diff-tree}\footnote{\url{https://git-scm.com/docs/git-diff-tree/}} command with rename detection enabled and the default similarity threshold of 50\% to accurately extract the Java files that differ between the two commits. This ensures that a renamed file is compared with its previous version instead of being counted as completely deleted and added again.

Next, we classify each changed Java file as either test or production code using the heuristic that was developed and validated in our previous study \citep{milanese2026}, which combines three rules: one matches the file path against test directories and naming conventions, the other two match the file content against imports of testing libraries and test-specific keywords, such as annotations and assertions. A file is classified as test code if it is in a standard test directory or matches at least two of the three rules, unless its path is in a build or dependency directory.

We remove single-line comments, block comments, and blank lines from both versions of each file, and trim lines to exclude changes to comments, whitespace, or indentation. Then, we compare the two stripped versions using the histogram algorithm from \texttt{git diff}\footnote{\url{https://git-scm.com/docs/git-diff/}}, which has been shown to be better suited for describing code changes~\citep{nugroho2020}. Finally, we count the changed lines following the \textsc{diffstat} algorithm \citep{dickey2026}. In each chunk of consecutive lines that are changed ($D$ deleted and $A$ added), the minimum number of lines $\min(D, A)$ are counted as modified, and the change size of the chunk is the maximum number of lines $\max(D, A)$. Summing the change sizes of the test and production file chunks yields the test change size $\Delta_{Test}$ and the production change size $\Delta_{Prod}$ of a \ac{PR}. We consider a \ac{PR} to be a \textit{test} \ac{PR} if $\Delta_{Test} > 0$, meaning that at least one non-comment, non-blank line of test code was changed.

We discard the 870 agent-supported and 1,315 human-authored \acp{PR} with $\Delta_{Test} + \Delta_{Prod} = 0$ because they only change comments or blank lines in Java files or have Java changes that cancel each other out between the pre- and post-merge commits.

\subsubsection{Repository Selection}
\label{sec:repository-selection}

Rather than selecting repositories based on popularity, we keep those with at least ten agent-supported and ten human-authored \acp{PR}. This criterion serves two purposes. First, it ensures that every repository contributes to both groups, which enables us to compare agent-supported and human-authored \acp{PR} within the same project while controlling for project-level factors, such as testing culture, domain, and build infrastructure. Second, it guarantees that there are sufficient \acp{PR} in each group to estimate how the difference between them varies across repositories.

Since steps \circledtext*[width=.9em,charshrink=.65]{3}, \circledtext*[width=.9em,charshrink=.65]{4}, and \circledtext*[width=.9em,charshrink=.65]{5} can only remove \acp{PR}, a repository can meet this criterion in the final dataset only if it met it after classification. Therefore, we apply it after each of these steps, as marked by the funnel icons in Figure \ref{fig:data-collection}. This avoids cloning and processing repositories that would be discarded anyway. Lastly, we discard repositories without any test \ac{PR}, since their \acp{PR} provide no test changes to characterize in terms of coverage or structure.

\subsubsection{Final Dataset}
\label{sec:data-characterization}

\begin{figure}[t]
    \centering
    \includegraphics[width=0.95\linewidth]{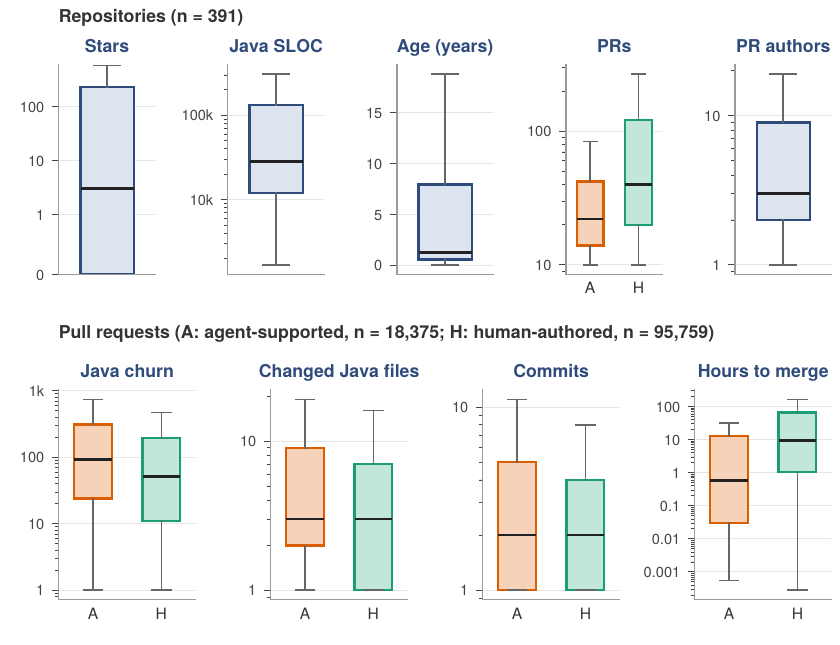}
    \caption{Distributions of repository and \ac{PR} characteristics in the final dataset. Outliers are not shown.}
    \label{fig:dataset-distributions}
\end{figure}

Our final dataset contains 114,134 \acp{PR} from 391 Java repositories, of which 18,375 are agent-supported and 95,759 human-authored. The distribution across agents is uneven: \textsc{Codex} (7,067), \textsc{Copilot} (5,764), and \textsc{Claude Code} (5,387) account for almost all agent-supported \acp{PR}, while \textsc{Cursor} (292) and \textsc{Devin} (9) contribute only a few, counting \acp{PR} that match several agents once for each. Moreover, agent-supported \acp{PR} become common only from mid-2025, and 40\% of them were merged in the last two months of our time window.

Figure~\ref{fig:dataset-distributions} shows the distributions of the main repository and \ac{PR} characteristics. The repositories are mostly small and young: the median repository has 3 stars, 28k Java SLOCs, and 3 \ac{PR} authors, and is 1.3 years old. Agent-supported \acp{PR} change more Java code than human-authored ones (median change size of 93 vs.\ 52 lines), although both groups share the same median number of changed Java files (3) and commits (2), and they are merged much faster (median of 0.6 vs.\ 9.7 hours). The complete dataset is available in our replication package~\citep{milanese2026r}.

\subsection{Experimental Procedure}
\label{sec:experimental-procedure}

All research questions share the same design. We compare agent-supported and human-authored \acp{PR}, or the test methods they change, within the same repository, using mixed-effects models in which each repository has its own baseline and its own difference between the two groups, and controlling for size. Unless stated otherwise, for each coefficient of \texttt{agentic} we report its exponential (an odds ratio or a rate ratio) with its 95\% Wald \ac{CI}, test whether it differs from 1 with a Wald \textit{z}-test \citep{agresti2006}, adjust \textit{p}-values with the Holm-Bonferroni method \citep{holm1979} within each family of tests, and consider a difference significant when its adjusted \textit{p}-value is below $\alpha = 0.05$ \citep{fisher1992}. We also report the predicted values for the two groups and the \ac{AME} of \texttt{agentic} \citep{arel-bundock2024}, the standard deviation $\hat{\tau}$ of the repository-specific differences, and evaluate model fit with simulation-based residual diagnostics \citep{hartig2026}. RQ$_1$ introduces this design in detail; for the other research questions, we describe the data they require, their metrics, and how their models and analyses depart from those of RQ$_1$.

\subsubsection{RQ$_1$: Frequency and Extent of Testing}
\label{sec:rq-1-experimental-procedure}

In RQ$_1$, we investigate whether agent-supported and human-authored \acp{PR} differ in how often and extensively they change test code.

\paragraph{Metrics.} We characterize testing based on four aspects: whether a \ac{PR} changes test code at all, whether a \ac{PR} changes only test code, and when a \ac{PR} changes both test and production code, how large the share of test changes is at the file and line levels.

\textbf{Testing Frequency.} To measure how often \acp{PR} change test code, we use the definition of test \ac{PR} of Section~\ref{sec:change-measurement}: the binary outcome $\mathit{IsTestPR}$ is 1 if $\Delta_{Test} > 0$ and 0 otherwise.

\textbf{Test-Only \acp{PR}.} A test \ac{PR} can change only test code, for example when a contributor adds tests for existing functionality, or both test and production code. Among test \acp{PR}, the binary outcome $\mathit{IsTestOnlyPR}$ is 1 if $\Delta_{Prod} = 0$ and 0 otherwise. We refer to test \acp{PR} that also change production code as \textit{mixed} \acp{PR}. This distinction is purely structural: whether the test and production changes of a mixed \ac{PR} co-evolve is covered in \hyperref[sec:rq-2-experimental-procedure]{RQ$_2$}.

\textbf{Testing Extent.} To measure the extent to which a \ac{PR} affects test code, we compute two ratios, at the file and line levels:
\begin{equation}
    R_{F} = \frac{\#\text{ files with } \Delta_{Test,f} > 0}{\#\text{ files with } \Delta_{Test,f} + \Delta_{Prod,f} > 0}, \qquad R_{\mathit{LOC}} = \frac{\Delta_{Test}}{\Delta_{Test} + \Delta_{Prod}}
\end{equation}
where $\Delta_{Test,f}$ and $\Delta_{Prod,f}$ are the test and production change sizes for file $f$. We only compute ratios for mixed \acp{PR} because \acp{PR} that do not change test code at all and test-only \acp{PR} are already covered by $\mathit{IsTestPR}$ and $\mathit{IsTestOnlyPR}$, respectively.

\paragraph{Statistical Model.} Because they share testing conventions, build infrastructure, and review practices, \acp{PR} from the same repository cannot be treated as completely independent observations. Moreover, agent-supported and human-authored \acp{PR} in our dataset are not distributed equally between repositories (Section~\ref{sec:data-characterization}). For these reasons, we analyze each outcome using \acp{GLMM} \citep{breslow1993}. \acp{GLMM} extend generalized linear models with random effects that capture variation between groups of observations \citep{stroup2024}, in our case, repositories.

We specify all four models as:

\begin{equation}
    \texttt{outcome} \sim \texttt{agentic} + \texttt{size} + (1 + \texttt{agentic} \mid \texttt{repository})
\end{equation}
where \texttt{agentic} is 1 for agent-supported \acp{PR} and 0 for human-authored ones. \texttt{size} is the log of the total Java change size of the \ac{PR} and is included as a control variable, since larger \acp{PR} are more likely to change test code and modify more files. The term $(1 + \texttt{agentic} \mid \texttt{repository})$ represents the random effects, allowing each repository to have its own baseline level of testing and difference between the two groups of \acp{PR}. Therefore, the coefficient of \texttt{agentic} measures the difference between agent-supported and human-authored \acp{PR} of the same size within the same repository, averaged across all repositories. Since this difference can vary between repositories, the model also estimates the spread of repository-specific differences around the average.

We model the two binary outcomes $\mathit{IsTestPR}$ and $\mathit{IsTestOnlyPR}$ with a Bernoulli distribution and the ratios $R_F$ and $R_{\mathit{LOC}}$ with a beta distribution \citep{ferrari2004}, which is suited to continuous values between 0 and 1. In both cases, we use the logit link function \citep{stroup2024}, and fit the models using the \textsc{glmmTMB} R package \citep{brooks2017}.

\paragraph{Statistical Analysis.} For each outcome, we measure the difference between agent-supported and human-authored \acp{PR} using the conditional odds ratio $OR = e^{\hat{\beta}}$ \citep{stroup2024}, where $\hat{\beta}$ is the fitted coefficient of \texttt{agentic}. For $\mathit{IsTestPR}$ and $\mathit{IsTestOnlyPR}$, the odds ratio compares the odds that an agent-supported \ac{PR} and a human-authored \ac{PR} of the same size in the same repository will change test code or only test code, respectively. For $R_F$ and $R_{\mathit{LOC}}$, the odds ratio compares the odds of their mean values. An odds ratio greater than 1 indicates higher values of the outcome for agent-supported \acp{PR}. Then, we compute the 95\% Wald \ac{CI} for each \citep{agresti2006}. To assess statistical significance, we test whether the odds ratios differ from 1 using a Wald \textit{z}-test \citep{agresti2006}. Since we test one hypothesis for each of the four outcomes, we adjust the resulting \textit{p}-values using the Holm-Bonferroni method \citep{holm1979} and consider a difference to be significant when its adjusted \textit{p}-value is below $\alpha = 0.05$ \citep{fisher1992}.

Since the same odds ratio corresponds to different changes in the outcome depending on its baseline value, we also express each difference on the original outcome scale through the \ac{AME} \citep{agresti2006} of \texttt{agentic}. For each outcome, we report the predicted mean for agent-supported and human-authored \acp{PR}, as well as the \ac{AME}. We compute the 95\% \acp{CI} of the \ac{AME} using the delta method with the \textsc{marginaleffects} R package \citep{arel-bundock2024}. To quantify how the difference between the two groups of \acp{PR} varies across repositories, we also report $\hat{\tau}$, the estimated standard deviation of the repository-specific deviations from the average coefficient of \texttt{agentic}. A $\hat{\tau}$ close to 0 indicates that the difference between the two groups is similar across all repositories, while larger values indicate greater variation from one repository to another.

Finally, we evaluate the fit of each model with simulation-based residual diagnostics using the \textsc{DHARMa} R package \citep{hartig2026}.

\subsubsection{RQ$_2$: Type and Context of Testing}
\label{sec:rq-2-experimental-procedure}

The definition of mixed \ac{PR} in \hyperref[sec:rq-1-experimental-procedure]{RQ$_1$} is purely structural, so its test changes are not necessarily related to its production changes. Here, we consider a test change to be co-evolving only if it is linked to production code changed by the same \ac{PR}~\citep{marsavina2014} (Section~\ref{sec2}), and ask whether mixed agent-supported and human-authored \acp{PR} co-evolve tests equally often, and whether the two groups differ in how often they add, modify, or delete tests in each development context.

\paragraph{Data Preparation.} Since this research question focuses on individual tests rather than files, the change measurement introduced in Section~\ref{sec:change-measurement} is insufficient to answer it. Therefore, we use \textsc{Hamster} \citep{pan2026} to determine which test methods and \textit{fixtures} (i.e., setup and teardown methods) each test \ac{PR} changes, and whether these are linked to the production code changes.

\textsc{Hamster} builds a model of each test class using the static analysis provided by \textsc{CLDK} \citep{krishna2025}, which resolves types and method calls from source code. To keep the analysis tractable, we run it on a \textit{slice} of each project consisting of the changed test and production files, the files they depend on and their superclasses, at both the pre- and post-merge commits. Within each changed test class, \textsc{Hamster} identifies test methods and fixtures, which we call \textit{test units}. For each test method, \textsc{Hamster} records the calls and assertions performed and the structural metrics we analyze in \hyperref[sec:rq-3-experimental-procedure]{RQ$_3$}. When a test method calls a \textit{helper method} (i.e., a non-test method in the same test class), \textsc{Hamster} adds the helper's calls and assertions to the test's model.

We match each unit across the two commits by its signature, using \textsc{RefactoringMiner} \citep{tsantalis2018, tsantalis2022} to detect renamed and moved units. A unit is \textit{added} or \textit{deleted} if it exists in only one version and is \textit{modified} if its text differs after removing comments and blank lines, as in Section~\ref{sec:change-measurement}. We also consider a unit modified when its text is unchanged but its \textsc{Hamster} model differs, since it then exercises a different behavior. Changed units that \textsc{Hamster} does not detect cannot be analyzed and are excluded.

A changed unit is \textit{linked} to the production changes of its \ac{PR} if it calls a production method that is added, deleted, or modified in the \ac{PR}, either directly or through a helper. Added units are linked through their post-merge version, deleted units through their pre-merge version, and modified units through either. Since \textsc{Hamster} resolves calls statically, the number of co-evolving \acp{PR} is a lower bound.

Of the 67,896 test \acp{PR} in our dataset, we successfully process 66,864 (98.48\%), 56,153 of which (83.98\%) change at least one test unit. The remaining \acp{PR} only change helper methods, fields, imports, or other test code outside of units.

\paragraph{Metrics.} Each changed unit has a task and a context. As in our previous study \citep{milanese2026}, we derive the task from \cite{pinto2012}: $\mathit{Add}$ if it is added, $\mathit{Mod}$ if it is modified, either directly or through a helper, and $\mathit{Del}$ if it is deleted. Units that are only renamed or moved have no task.

The context of a unit depends on its \ac{PR} and on its linkage:

\begin{itemize}
    \item Test-Focused Context ($\mathit{TFC}$): units of a test-only \ac{PR}, which cannot be linked since the \ac{PR} changes no production code.
    \item Co-Evolution Context ($\mathit{COC}$): units of a mixed \ac{PR} that are linked to production code changes.
    \item Test-Focused Context in Mixed \acp{PR} ($\mathit{TFM}$): units of a mixed \ac{PR} for which no link to its production code changes is detected.
\end{itemize}

\textbf{Co-Evolution.} On mixed \acp{PR}, the binary outcome $\mathit{IsCoEvolutionPR}$ is 1 if at least one changed unit of the \ac{PR} is linked to its production changes, and 0 otherwise. It includes units modified through a helper, since a test whose helper now calls changed production code co-evolves with it.

\textbf{Types of Test Changes.} For each task $t \in \{\mathit{Add},\mathit{Mod},\mathit{Del}\}$ and context $k \in \{\mathit{TFC},\mathit{COC},\mathit{TFM}\}$, the binary outcome $t_k$ is 1 if the \ac{PR} has at least one changed unit in context $k$ whose task is $t$, and 0 otherwise. $\mathit{TFC}$ is defined on test-only \acp{PR}, and $\mathit{COC}$ and $\mathit{TFM}$ on mixed \acp{PR}. As in our previous study \citep{milanese2026}, these outcomes are not mutually exclusive: a mixed \ac{PR} that adds a linked test and modifies an unlinked one counts for both $\mathit{Add}_{\mathit{COC}}$ and $\mathit{Mod}_{\mathit{TFM}}$.

\paragraph{Statistical Model.} We model $\mathit{IsCoEvolutionPR}$ on mixed \acp{PR} using the same specification as in \hyperref[sec:rq-1-experimental-procedure]{RQ$_1$}. For the nine outcomes $t_k$, we fit one model per task, each covering all three contexts. In each task model, an observation is a \ac{PR} in one context, and its outcome is whether the PR performs the task in that context. A test-only \ac{PR} only has $\mathit{TFC}$ units, so it contributes one observation: in the model for $\mathit{Add}$, for example, its outcome is $\mathit{Add_{\mathit{TFC}}}$. A mixed \ac{PR} can have both linked and unlinked units, so it contributes two observations, one for $\mathit{COC}$ and one for $\mathit{TFM}$.

We specify the three models as:

\begin{equation}
\begin{split}
    \texttt{outcome} \sim\ & \texttt{0} + \texttt{context} + \texttt{context:agentic} + \texttt{context:size} \\
    & + (1 + \texttt{agentic} \mid \texttt{repository}) + (1 \mid \texttt{repository:context})
\end{split}
\end{equation}
where \texttt{agentic} and \texttt{size} are defined as in \hyperref[sec:rq-1-experimental-procedure]{RQ$_1$}, and \texttt{context} identifies the context of the observation. The leading \texttt{0} removes the common intercept, so the fixed effects give each context its own intercept, \texttt{agentic} coefficient, and \texttt{size} coefficient. Therefore, the coefficient of \texttt{agentic} in context $k$ measures the difference between agent-supported and human-authored \acp{PR} of the same size within the same repository and context. The random effects differ from those of \hyperref[sec:rq-1-experimental-procedure]{RQ$_1$} in two ways. First, the repository-specific difference between the two groups of \acp{PR}, $(1 + \texttt{agentic} \mid \texttt{repository})$, is shared by the three contexts. Second, the term $(1 \mid \texttt{repository:context})$ allows the baseline of each repository to differ between contexts, since projects differ in how much test-focused work they perform relative to co-evolution. Fitting the three contexts in one model also allows us to test whether the difference between agent-supported and human-authored \acp{PR} depends on the context.

As in RQ$_1$, outcomes follow a Bernoulli distribution with the logit link.

\paragraph{Statistical Analysis.} We analyze $\mathit{IsCoEvolutionPR}$ and the nine context-specific coefficients of \texttt{agentic} as one family of tests, and compute the \ac{AME}, in percentage points, for each context of the task models.

To assess whether the difference between the two groups depends on the context, we test for each task the null hypothesis $\hat{\beta}_{\mathit{TFC}} = \hat{\beta}_{\mathit{COC}} = \hat{\beta}_{\mathit{TFM}}$ with a Wald $\chi^2$ test \citep{agresti2006}, adjusting the three \textit{p}-values with the Holm-Bonferroni method. To show where differences arise, we also report, without testing it, the ratio between the odds ratios of each pair of contexts, where a ratio of 1 indicates that the difference between the two groups is the same in both contexts.

\subsubsection{RQ$_3$: Test Structure}
\label{sec:rq-3-experimental-procedure}

In RQ$_3$, we investigate whether the test methods added and modified by agent-supported and human-authored \acp{PR} differ in structure in terms of oracles, assertions, conditional logic, call-assertion structure, mocking, scope, and length. While \hyperref[sec:rq-2-experimental-procedure]{RQ$_2$} covers the types of test changes made by the two groups of PRs, RQ$_3$ characterizes the tests themselves.

\paragraph{Data Preparation.} We reuse the test units identified by \textsc{Hamster} in \hyperref[sec:rq-2-experimental-procedure]{RQ$_2$} and analyze the test methods added or modified by a \ac{PR}, excluding fixtures and test methods modified only through a helper. Our primary focus is on added test methods because they are entirely written within the \ac{PR}. Modified test methods, on the other hand, mostly reflect the original authors' work, so we only analyze the changes contributed by the \ac{PR} itself.

For each test method, the \textsc{Hamster} model records its annotations, size, cyclomatic complexity, created objects, mocking usage, and focal classes and methods. It also splits the body of the test method into call-assertion sequences and classifies each assertion by type \citep{pan2026}.

Of the 443,072 test methods added by test \acp{PR}, \textsc{Hamster} locates 397,021 (89.6\%), from 35,801 \acp{PR}: 130,067 from 7,070 agent-supported and 266,954 from 28,731 human-authored \acp{PR}. Of the 429,281 test methods they modify, it locates both versions of 394,610 (91.9\%), from 30,475 \acp{PR}: 57,858 from 4,115 agent-supported and 336,752 from 26,360 human-authored \acp{PR}.

\paragraph{Metrics.} We categorize added test methods based on five dimensions: oracle, conditional logic, call-assertion structure, mocking, and scope, and also report their length. Since many tests lack conditional logic and mocks, we define a binary outcome that indicates whether a test method has these characteristics, as well as a count that measures how much of each characteristic the test method has. This allows the two aspects to differ between agent-supported and human-authored \acp{PR}.

For modified test methods, instead, we measure how the \ac{PR} changes their assertions and whether it removes their oracle, and define corresponding binary outcomes.

\textbf{Oracle.} A test method $m$ has an \textit{explicit oracle}, i.e., $\mathit{HasOracle}(m) = 1$, if it contains a check that can make it fail when the behavior it exercises is incorrect: $a(m) + v(m) > 0$ or $\mathit{expects}(m)$, where $a(m)$ is the number of assertions in its own body, $v(m)$ the number of verification calls of mocking frameworks in its own body (e.g., \textsc{Mockito}'s \texttt{verify}), and $\mathit{expects}(m)$ indicates an expected exception declared in its annotation. We do not count the assertions of the helper methods that $m$ calls, so a test method that delegates all of its checks to helpers has no explicit oracle by this definition. In assertion chains (e.g., \texttt{assertThat(x).isEqualTo(y)}), the call that opens the chain is not counted, so a chain with a single check counts as one assertion. Among test methods with at least one assertion, the count $\mathit{Assertions}$ is $a(m)$.

For a modified test method, we compare its assertions before and after the \ac{PR}. An assertion counts as removed if its code no longer occurs in the test method after the \ac{PR}, and as added if it did not occur before, so editing an assertion counts as one removal and one addition. $\mathit{RemovesAssertions}$ and $\mathit{AddsAssertions}$ indicate that the \ac{PR} removes or adds at least one assertion, respectively. We consider removals and additions separately, since a \ac{PR} can replace one assertion with another. These outcomes describe syntactic changes: they do not measure whether a change strengthens or weakens what the test checks, since an added assertion may be redundant and a rewritten one may be stronger. $\mathit{LosesOracle}$ indicates instead that the test method has an explicit oracle before the \ac{PR} but not after it, the most direct syntactic form of weakening.

\textbf{Conditional Logic.} The decision points of a test method are the points at which its execution can branch, and their number equals the cyclomatic complexity of the method minus one \citep{mccabe1976}. We define two variables: the binary outcome $\mathit{HasCondLogic}$ indicates whether a test method has at least one decision point, and the count $\mathit{DecisionPoints}$ is the number of decision points in test methods that have at least one.

\textbf{Call-Assertion Structure.} \textsc{Hamster} divides each test method into \textit{call-assertion sequences}, each consisting of calls followed by assertions that check the results of the calls \citep{pan2026}. Since \textsc{Hamster} adds the calls and assertions of the helper methods that a test method calls to its model, sequences include those of its helpers. The count outcome $\mathit{Sequences}$ is the number of call-assertion sequences in test methods with at least one.

\textbf{Mocking.} We count calls that create mocks or spies within a test method's body (e.g., \textsc{Mockito}'s \texttt{mock()} and \texttt{spy()}). The binary outcome $\mathit{CreatesMocks}$ indicates whether a test method creates at least one mock, and the count $\mathit{Mocks}$ is the number of mocks created by test methods that create at least one. Mocks that are declared as annotated fields of the test class are shared by all of its test methods and are not attributed to any method.

\textbf{Scope.} \textsc{Hamster} identifies the \textit{focal methods} of a test method, i.e., the production methods under test, and, by extension, its \textit{focal classes}. We only measure scope on test methods with at least one resolved focal class. On these, the count $\mathit{FocalClasses}$ is the number of focal classes, and the count $\mathit{FocalMethods}$ is the number of focal methods, which can be zero when a test method only creates instances of its focal class, for example to check that a constructor initializes its fields, or only calls trivial methods such as getters, which Hamster does not consider to be focal methods \citep{pan2026}.

\textbf{Length.} $\mathit{Length}$ is the number of lines of a test method, excluding comments and blank lines as in Section~\ref{sec:change-measurement}.

\paragraph{Statistical Model.} We fit one \ac{GLMM} per outcome, extending the specification of \hyperref[sec:rq-1-experimental-procedure]{RQ$_1$} in two ways:

\begin{equation}
    \texttt{outcome} \sim \texttt{agentic} + \texttt{size} + (1 + \texttt{agentic} \mid \texttt{repository}) + (1 \mid \texttt{pr})
\end{equation}
First, observations are now test methods, and those of the same \ac{PR} cannot be treated as independent. Without accounting for this, a \ac{PR} that adds many similar tests would carry as much weight as many \acp{PR} adding one test each, so the term $(1 \mid \texttt{pr})$ allows each \ac{PR} to have its own baseline within its repository. When the variation between \acp{PR} cannot be estimated, i.e., its standard deviation exceeds 5 on the link scale, we refit the model without this term, which is the case for $\mathit{CreatesMocks}$ and $\mathit{LosesOracle}$. Second, \texttt{size} is now the log of the size of the test method: its length for added test methods, its length including the helper methods it calls for $\mathit{Sequences}$, consistently with how sequences are counted, and the number of lines that the \ac{PR} changes in it for modified test methods. We center \texttt{size} within the test methods of each outcome. As a result, the coefficient of \texttt{agentic} measures the difference between agent-supported and human-authored test methods of the same size within the same repository.

We model the binary outcomes with a Bernoulli distribution and the logit link. Since the counts $\mathit{Assertions}$, $\mathit{DecisionPoints}$, $\mathit{Sequences}$, $\mathit{Mocks}$, and $\mathit{FocalClasses}$ are only measured on test methods with at least one, we model them with a zero-truncated negative binomial distribution, and $\mathit{FocalMethods}$ with a negative binomial distribution, both with the log link \citep{hilbe2011}. We model the log of $\mathit{Length}$ with a Gaussian distribution and the same random effects, but without \texttt{size}, so that its coefficient compares the lengths of the two groups. Again, we fit the models using \texttt{glmmTMB} \citep{brooks2017}.

\paragraph{Development Context.} To assess whether the difference between the two groups depends on why tests are added, we assign each added test method to one of the contexts of \hyperref[sec:rq-2-experimental-procedure]{RQ$_2$}: $\mathit{TFC}$ if its \ac{PR} is test-only, $\mathit{COC}$ if it is linked to the production changes of a mixed \ac{PR}, and $\mathit{TFM}$ otherwise. For each outcome on added test methods, we refit the model with the fixed effects \texttt{0 + context + context:agentic + context:size}, centering \texttt{size} within each context, and add the term $(1 \mid \texttt{repository:context})$, as in \hyperref[sec:rq-2-experimental-procedure]{RQ$_2$}. Unlike in \hyperref[sec:rq-2-experimental-procedure]{RQ$_2$}, each test method belongs to a single context, and the term $(1 \mid \texttt{pr})$ accounts for the test methods of a mixed \ac{PR} that fall in both $\mathit{COC}$ and $\mathit{TFM}$.

\paragraph{Statistical Analysis.} We measure the difference between agent-supported and human-authored test methods for each outcome with the conditional odds ratio for binary outcomes, the rate ratio $RR$ for counts \citep{hilbe2011}, and the ratio of geometric means $\mathit{GMR}$ for $\mathit{Length}$, all computed as $e^{\hat{\beta}}$ where ${\hat{\beta}}$ is the coefficient of \texttt{agentic}. For zero-truncated counts, $e^{\hat{\beta}}$ is the ratio between the means of the untruncated distributions, which differs from the ratio between the mean counts of test methods with at least one, so we also compare these means directly through predictions on the response scale. We adjust \textit{p}-values separately for the nine outcomes on added test methods and for the three outcomes on modified test methods. Since $\mathit{Length}$ describes the test methods rather than testing a hypothesis, we report its \textit{p}-value without adjustment. As we did not set equivalence margins in advance, we interpret non-significant differences through their \acp{CI}, i.e., as bounds on the differences that are compatible with the data, rather than as evidence of equivalence.

To assess whether the difference depends on the context, we test for each outcome the null hypothesis that the coefficient of \texttt{agentic} is the same in all three contexts with a Wald $\chi^2$ test \citep{agresti2006}, adjusting \textit{p}-values with the Holm-Bonferroni method across the nine outcomes \citep{holm1979}, and report the context-specific ratios with their 95\% Wald \acp{CI}.

As in \hyperref[sec:rq-1-experimental-procedure]{RQ$_1$}, we predict each test method both as human-authored and as agent-supported with its own size and all random effects at zero, i.e., for a typical \ac{PR} in a typical repository, and average the predictions. For counts, predictions are on the response scale, i.e., the mean count among test methods with at least one. For binary outcomes, we also report the share of the latent variance that lies between repositories and \acp{PR}.

\subsubsection{RQ$_4$: Test Smells}
\label{sec:rq-4-experimental-procedure}

In RQ$_4$, we investigate whether agent-supported and human-authored \acp{PR} differ in how they affect the evolution of test smells.

\paragraph{Data Preparation.} For each test-related \ac{PR}, we analyze the test suite immediately before the contribution (\emph{pre-PR}) and after it (\emph{post-PR}) using \textsc{JNose}~\citep{virginio2020jnose}, a static-analysis-based test smell detector specifically designed for Java. In our previous study \citep{milanese2026}, we used the language-independent \textsc{AromaDr} detector~\citep{silva2025aromadr}. Since the present study focuses exclusively on Java, we use \textsc{JNose} because it provides a broader catalog of 21 test smells, therefore, 11 more categories than \textsc{AromaDr}, and reports individual occurrences together with their source-code location. The complete catalog of the 21 categories is described by \citet{virginio2020jnose}.

We run \textsc{JNose-Core} independently on the complete pre- and post-PR snapshots. Test code unaffected by the contribution appears in both revisions and therefore cancels out when computing the change in smell occurrences. The fine-grained locations reported by \textsc{JNose} allow us to retain the results at two complementary levels: the complete \ac{PR} and the individual test method.

We also distinguish between Addition and Modification contributions using the classification available in \texttt{pull\_request\_analysis}. Addition \acp{PR} do not alter pre-existing test logic, whereas Modification \acp{PR} change existing tests. We analyze the two groups separately because creating new test code and maintaining existing tests may exhibit different test-quality dynamics.

\paragraph{Metrics.} For each smell category $s$, we measure its evolution through the \emph{Smell Delta}:

\begin{equation}
    \Delta_s = N^{post}_s - N^{pre}_s ,
\end{equation}

where $N^{pre}_s$ and $N^{post}_s$ are the numbers of occurrences detected before and after the contribution. Positive values indicate a net increase in the smell, negative values indicate a net reduction, and zero indicates no net change. Using the delta rather than the absolute post-PR count prevents pre-existing smell occurrences from being attributed to the contribution.

We compute this metric at two granularities. At the \ac{PR} level, smell occurrences are aggregated over the complete contribution, capturing its overall effect on the test suite. At the method level, the delta is computed for each individual test method, allowing us to determine whether differences between agent-supported and human-authored contributions remain observable at a finer granularity. For newly added methods, the pre-PR smell count is zero, whereas for removed methods the post-PR count is zero.

In addition to the category-specific deltas, we compute the \emph{Total Smell Delta}, obtained by considering all detected smell occurrences together. Although \textsc{JNose} supports 21 categories, one category exhibits no variation in our data; therefore, the statistical analysis covers the remaining 20 categories and the aggregate outcome.

\paragraph{Statistical Model.} As in the previous research questions, observations originating from the same repository cannot be assumed to be independent. We therefore fit linear mixed-effects models using the same repository-level random-effects structure introduced in \hyperref[sec:rq-1-experimental-procedure]{RQ$_1$}, allowing both the baseline and the difference between agent-supported and human-authored contributions to vary across repositories.

At the \ac{PR} level, we specify the model as:

\begin{equation}
\begin{split}
\texttt{delta} \sim\;&
\texttt{agentic}
+ \log(\texttt{added\_change\_size}+1) \\
&+ \log(\texttt{modified\_change\_size}+1)
+ (1+\texttt{agentic}\mid\texttt{repository}).
\end{split}
\end{equation}

The two change size terms control for the amount of newly added and modified test code, since larger contributions have more opportunities to introduce or remove smell occurrences. For Addition \acp{PR}, \texttt{modified\_change\_size} is omitted because pre-existing test logic is not modified.

At the method level, we extend the same model with $(1\mid\texttt{pr})$, following the approach adopted in \hyperref[sec:rq-3-experimental-procedure]{RQ$_3$}, since multiple test methods belonging to the same \ac{PR} are not independent. The coefficient of \texttt{agentic} therefore estimates the average difference in Smell Delta between agent-supported and human-authored contributions with comparable change size within repositories.

\paragraph{Statistical Analysis.} For each model, our primary quantity of interest is the coefficient of \texttt{agentic}. A positive coefficient indicates that agent-supported contributions exhibit a larger Smell Delta than human-authored contributions, whereas a negative coefficient indicates a smaller Smell Delta. Since the outcome represents a net difference, the coefficient captures the overall evolution of the smell and should not necessarily be interpreted as the number of smell instances directly introduced by the contribution.

We report the coefficient with its 95\% \ac{CI} and assess statistical significance at $\alpha=0.05$. Because the same comparison is repeated across 20 smell categories, we adjust the corresponding $p$-values using the Holm-Bonferroni method~\citep{holm1979}. The correction is applied separately for each combination of analysis level (PR or method) and contribution type (Addition or Modification), while the Total Smell Delta is analyzed as a separate aggregate outcome.

\subsubsection{RQ$_5$: Code Coverage}
\label{sec:rq-5-experimental-procedure}

In RQ$_5$, we investigate whether agent-supported and human-authored \acp{PR} differ in how much code coverage they introduce and how well the test changes they make keep pace with the production code they modify. Unlike the other research questions, which analyze source code, RQ$_5$ requires building each project and executing its test suite before and after every \ac{PR}. For this reason, it only covers the test \acp{PR} of the repositories that we can build automatically.

\paragraph{Data Preparation.} For each repository, we try combinations of build tools (\textsc{Maven} or \textsc{Gradle}, using the project's wrapper if available) and Java versions supported by the project until we find one under which it builds successfully. Of the 391 repositories in our dataset, 150 pass this step.

For each test \ac{PR} of these repositories, we run the test suite twice: once at the pre-merge commit and once at the post-merge commit identified in Section~\ref{sec:pre-post-merge-commit}. This ensures that the two runs differ only by the changes introduced by the \ac{PR} itself. Each run occurs in a \textsc{Docker}\footnote{\url{https://www.docker.com/}} container with time limits of 15 minutes for the build and 30 minutes for the tests. To record which lines of code each test executes, we run the \textsc{JUnit} 4 and \textsc{JUnit} 5 tests for each module with the \textsc{JaCoCo} agent and a custom listener that saves and resets coverage data at the beginning and end of each test method. To identify the test methods that each \ac{PR} adds, modifies, or deletes, i.e., the \textit{tests contributed in the \ac{PR}}, we reuse the \textsc{Hamster} models built in \hyperref[sec:rq-2-experimental-procedure]{RQ$_2$}. Since \textsc{Hamster} does not always identify the contributed tests correctly, for the metrics that depend on them we exclude the \acp{PR} where more than 10\% of the contributed tests are missing from the run in which they should appear, or where tests present in only one of the two runs are not identified as contributed. We also record every coverable line of every class so lines that no test executes are counted as well.

We only analyze a \ac{PR} when both runs complete and we can verify that they were executed at the expected commits. This results in 3,619 test \acp{PR} (1,552 agent-supported and 2,067 human-authored) from 95 repositories.

\paragraph{Metrics.} All of our metrics measure line coverage, i.e., the proportion of executable lines covered by at least one test, but they differ in which lines and tests they consider. Given a set of classes $C$ and a set of tests $T$, we define coverage at commit $c \in \{\mathit{pre}, \mathit{post}\}$ of a \ac{PR} as:

\begin{equation}
    \mathit{Cov}_c(C, T) = \frac{|\{\ell \in L_c(C) : \ell \text{ is executed by some } t \in T_c\}|}{|L_c(C)|}
\end{equation}
where $L_c(C)$ is the set of coverable lines of the classes in $C$, and $T_c$ is the set of tests in $T$ executed in the run at commit $c$.

Let $K$ be all the classes and $A$ all the tests of the project, $T^{\mathit{add}}$, $T^{\mathit{mod}}$, and $T^{\mathit{del}}$ the tests that the \ac{PR} adds, modifies, and deletes, $P$ the classes of the production files that it changes, and $X$ the classes executed by at least one test contributed in the \ac{PR}. We define the five metrics as:

\textbf{Codebase Coverage.} $\mathit{Cov}_c(K, A)$ is the coverage of all the classes of the project by its whole test suite. We compute it on all test \acp{PR}, both test-only and mixed, with at least one coverable line at both commits.

\textbf{Affected-Class Coverage.} $\mathit{Cov}_c(X \cup P, A)$ is the coverage, by the whole test suite, of the classes that a \ac{PR} contributes to, either because it changes their production code or because the tests contributed in the \ac{PR} exercise them. We compute it on all test \acp{PR}, both test-only and mixed, with at least one coverable line at the post-merge commit.

\textbf{Contributed-Test Coverage.} $\mathit{Cov}_{\mathit{pre}}(X, T^{\mathit{del}} \cup T^{\mathit{mod}})$ and $\mathit{Cov}_{\mathit{post}}(X, T^{\mathit{add}} \cup T^{\mathit{mod}})$ are the coverage of the classes executed by the tests contributed in the \ac{PR} by these tests alone, i.e., the deleted and modified tests at the pre-merge commit, and the added and modified tests at the post-merge commit. We compute it on the test \acp{PR}, both test-only and mixed, that add or modify at least one test, with at least one coverable line at the post-merge commit.

\textbf{Production-Change Coverage.} $\mathit{Cov}_c(P, A)$ is the coverage of the production classes changed in the \ac{PR} by the whole test suite. Since test-only PRs do not change production code, we compute it only on mixed PRs, with at least one coverable line at the post-merge commit.

\textbf{New-Line Coverage.} It is defined on the set $N$ of coverable lines that the \ac{PR} adds to production files. Let $N_A \subseteq N$ be the lines executed by the whole test suite at the post-merge side, and $N_T \subseteq N_A$ those executed by at least one test added or modified in the \ac{PR}. We measure the share of new lines that are covered, $|N_A| \,/\, |N|$, and, among these, the share covered by the tests contributed in the \ac{PR}, $|N_T| \,/\, |N_A|$. We compute both shares only on mixed PRs, since test-only PRs add no production lines: the first on those that add at least one coverable production line, and the second on those with at least one covered new line.

\paragraph{Statistical Model.} For the \textit{Codebase Coverage} metric, we fit two logistic regression models, one for whether it increases and one for whether it decreases between the pre- and post-merge commits. As it is defined on both test-only and mixed \acp{PR}, we use the specification of \hyperref[sec:rq-2-experimental-procedure]{RQ$_2$}, with the type of \ac{PR} in place of the context:

\begin{equation}
\begin{split}
    \texttt{outcome} \sim\;&
    0 + \texttt{type} + \texttt{type:agentic} + \texttt{type:size} \\
    &+ (1 + \texttt{agentic} \mid \texttt{repository}) + (1 \mid \texttt{repository:type})
\end{split}
\end{equation}
Each type of \ac{PR} has its own intercept and coefficients, and the term $(1 \mid \texttt{repository:type})$ allows the baseline of each repository to differ between types.

We model the other metrics as the number of covered lines out of the number of coverable lines, using a beta-binomial distribution with the logit link \citep{harrison2015}. Unlike the beta distribution we use for the ratios in \hyperref[sec:rq-1-experimental-procedure]{RQ$_1$}, it allows coverage to be exactly 0 or 1, which is common in our data, and it takes the number of coverable lines into account, so that a \ac{PR} with 2 coverable lines weighs less than one with 200.

\textit{Affected-Class Coverage} and \textit{Contributed-Test Coverage} are also defined on both types of \ac{PR}, so we use the same specification as for \textit{Codebase Coverage}, with the number of covered lines at the post-merge commit as the outcome. For \textit{Contributed-Test Coverage}, we only model the post-merge value, since the pre-merge one is 0 by construction for \acp{PR} that only add tests.

For \textit{Affected-Class Coverage} we add two terms to the model, \texttt{baseline} and \texttt{no\_baseline}, which account for the coverage before the \ac{PR}. We need these because the coverage of a class after a \ac{PR} largely depends on its coverage before it, and agent-supported and human-authored \acp{PR} may change code that was already covered to a different extent. \texttt{baseline} is the pre-merge coverage of the same classes, on the logit scale, so that the coefficient of \texttt{agentic} compares \acp{PR} that start from the same coverage. When all the classes are new, there is no pre-merge coverage: rather than dropping these \acp{PR}, we flag them with \texttt{no\_baseline}. We prefer this to modeling the change in coverage, which can lead to opposite conclusions when the two groups start from different coverage.

Finally, \textit{Production-Change Coverage} and \textit{New-Line Coverage} are only defined on mixed \acp{PR}, so we use the same specification as in \hyperref[sec:rq-1-experimental-procedure]{RQ$_1$}, adding the baseline terms we introduced for \textit{Production-Change Coverage}:

\begin{equation}
\begin{split}
    \texttt{coverage} \sim\;& \texttt{agentic} + \texttt{size} + \texttt{baseline} + \texttt{no\_baseline} \\
    &+ (1 + \texttt{agentic} \mid \texttt{repository})
\end{split}
\end{equation}

When a model does not converge, we simplify its random effects, first removing the correlation between the repository intercept and slope and then the slope, each time with and without $(1 \mid \texttt{repository}:\texttt{type})$, and report the structure used for each metric. All models are fitted with the \textsc{glmmTMB} R package \citep{brooks2017}.

\paragraph{Statistical Analysis.} For each metric and type of \ac{PR}, we compute the conditional odds ratio of \texttt{agentic}. Adding a covariate strongly associated with the outcome changes an odds ratio even without confounding \citep{greenland1999}, so the odds ratios of \textit{Affected-Class} and \textit{Production-Change Coverage}, which are adjusted for the baseline, are not directly comparable with those of the other metrics. We therefore use the \ac{AME} of \texttt{agentic} as our main effect size, i.e., the difference in mean predicted coverage, in percentage points, between all agent-supported and all human-authored \acp{PR}, with repository effects set to zero, i.e., for a typical repository, as in \hyperref[sec:rq-1-experimental-procedure]{RQ$_1$}. We compute its 95\% \ac{CI} from 1,000 draws of the fixed effects from their estimated sampling distribution.

We adjust \textit{p}-values within the family of the six main tests. We test the share of new lines covered by the tests contributed in the \ac{PR} ($|N_T| \,/\, |N_A|$) separately, since it concerns which tests cover new code rather than how much of it is covered. For \textit{Affected-Class} and \textit{Contributed-Test Coverage}, we also test whether the coefficient of \texttt{agentic} is the same in test-only and mixed \acp{PR} with a Wald $\chi^2$ test \citep{agresti2006}, adjusting the two \textit{p}-values with the Holm-Bonferroni method, and report the ratio between the odds ratios of the two types. We do not adjust the tests of \textit{Codebase Coverage}, which only provides context.

\section{Results}\label{sec:results}

This section reports the results of each research question. For each, we first summarize the main differences between the two groups, then show how they depend on the development context, on the size of the change, and on the repository, and close with the model fit and a summary answer. Unless stated otherwise, predicted values and differences refer to a typical repository, i.e., with repository effects set to zero, and $p_{adj}$ denotes \textit{p}-values adjusted with the Holm-Bonferroni method within their family.

\subsection{RQ$_1$: Frequency and Extent of Testing}
\label{sec:rq-1-results}

Within the same repositories, agent-supported \acp{PR} change test code more often and more extensively than human-authored \acp{PR} of the same size. We first summarize these differences, and then show why pooled comparisons hide them, how they translate into practice for \acp{PR} of different sizes, and how they vary between repositories.

\paragraph{Agent-supported \acp{PR} test more often and more extensively.} Table~\ref{tab:rq1-models} reports the difference between the two groups on all four outcomes. All four odds ratios are above 1 and remain significant after the Holm-Bonferroni correction.

\begin{table}[t]
    \centering
    \caption{Difference between agent-supported and human-authored \acp{PR} within the same repository. The \textit{Human} and \textit{Agent} columns report the predicted value in a typical repository.}
    \label{tab:rq1-models}
    \small
    \setlength{\tabcolsep}{3.5pt}
    \begin{tabular}{lrrrrrr}
    \toprule
    Outcome & OR (95\% CI) & $p_{adj}$ & Human & Agent & AME (95\% CI) & $\hat{\tau}$ \\
    \midrule
    $\mathit{IsTestPR}$     & 1.65 (1.46, 1.88) & $<0.001$ & 45.8\% & 55.0\% & $+$9.2 (6.9, 11.5) & 0.98 \\
    $\mathit{IsTestOnlyPR}$ & 1.35 (1.12, 1.62) & $0.002$  & 13.4\% & 16.3\% & $+$2.9 (1.0, 4.7)  & 1.10 \\
    \addlinespace
    $R_F$                   & 1.12 (1.08, 1.16) & $<0.001$ & 33.8\% & 36.2\% & $+$2.5 (1.7, 3.3)  & 0.21 \\
    $R_{\mathit{LOC}}$      & 1.26 (1.20, 1.33) & $<0.001$ & 44.0\% & 49.8\% & $+$5.8 (4.5, 7.1)  & 0.34 \\
    \bottomrule
    \end{tabular}
\end{table}

In a typical repository, 55.0\% of agent-supported \acp{PR} change test code, against 45.8\% of human-authored ones ($+$9.2 pp, $OR$ 1.65). Among test \acp{PR}, agent-supported ones change only test code slightly more often ($+$2.9 pp, $OR$ 1.35), which is the smallest and least precise of the four effects. When a \ac{PR} changes both test and production code, agent-supported \acp{PR} devote a larger share of their changes to tests: $+$5.8 pp of the changed lines ($OR$ 1.26), but only $+$2.5 pp of the changed files ($OR$ 1.12). This suggests that agent-supported \acp{PR} change more test code within the test files they touch, rather than touching more test files.

\paragraph{Pooled comparisons hide these differences.} When all \acp{PR} are pooled, without accounting for repositories or size, the two groups look similar (Table~\ref{tab:rq1-descriptives}). Agent-supported and human-authored \acp{PR} change test code about equally often (60.1\% vs.\ 59.4\%), and human-authored \acp{PR} score slightly higher on the other three outcomes.

\begin{table}[t]
    \centering
    \caption{Pooled comparison of agent-supported and human-authored \acp{PR}.}
    \label{tab:rq1-descriptives}
    \small
    \begin{tabular}{lrr}
    \toprule
    Outcome & Agent & Human \\
    \midrule
    \quad $\mathit{IsTestPR}$, share of \acp{PR} & 60.1\% & 59.4\% \\
    \addlinespace
    \quad $\mathit{IsTestOnlyPR}$, share of test \acp{PR} & 15.2\% & 16.6\% \\
    \addlinespace
    \quad $R_F$, mean (median) & 37.3\% (37.0\%) & 39.3\% (40.0\%) \\
    \quad $R_{\mathit{LOC}}$, mean (median) & 51.7\% (52.8\%) & 52.7\% (54.5\%) \\
    \bottomrule
    \end{tabular}
\end{table}

The pooled comparison differs from the within-repository one mainly because the two groups come from different repositories, and agent-supported \acp{PR} are concentrated in repositories where \acp{PR} change test code less often. Comparing the raw shares within each repository, agent-supported \acp{PR} change test code more often in 272 of the 391 repositories and less often in 114. The same holds for the extent of testing. For test-only \acp{PR}, size contributes as well, as agent-supported test \acp{PR} are larger (median 161 vs.\ 104 lines), and larger test \acp{PR} less often change only test code.

This shows that comparisons of agent-supported and human-authored \acp{PR} that pool repositories, including those in our previous study \citep{milanese2026}, can largely reflect which projects the \acp{PR} come from.

\begin{figure}[t]
\centering
\includegraphics[width=\textwidth]{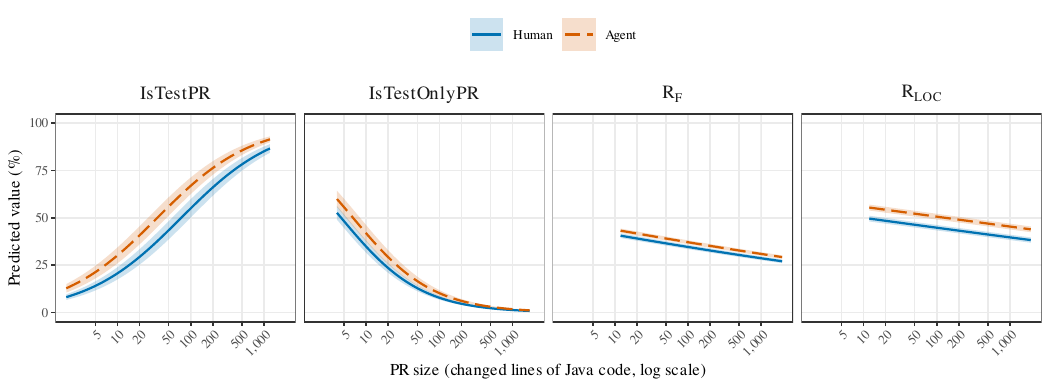}
\caption{Predicted outcomes for agent-supported and human-authored \acp{PR} in a typical repository, by size of the \ac{PR}}
\label{fig:rq1-size-curves}
\end{figure}

\paragraph{In practice, the difference is most visible in \acp{PR} of a few dozen lines.} Figure~\ref{fig:rq1-size-curves} shows the predicted outcomes for both groups in a typical repository across sizes. Since the models assume the same odds ratio at every size, the curves show how this one difference translates into percentages for \acp{PR} of different sizes. The vertical distance between the two curves is the difference in percentage points.

For testing frequency, the gap is widest for \acp{PR} of a few dozen lines, where human-authored \acp{PR} change test code in about half of the cases or fewer. At 50 lines, 56.4\% of agent-supported \acp{PR} change test code, against 43.8\% of human-authored ones ($+$12.5 pp). The gap narrows for very small \acp{PR}, which rarely change test code in either group (30.4\% vs.\ 20.9\% at 10 lines), and for very large ones, which almost always do (90.6\% vs.\ 85.3\% at 1,000 lines). The gap for test-only \acp{PR} instead shrinks steadily with size, because they become rare: from $+$7.0 pp at 10 lines to $+$3.7 pp at 50 lines and less than 1 pp at 1,000 lines. For the two shares, which stay between about 25\% and 60\% at all sizes, the gap is nearly constant, at about $+$5.8 pp of the changed lines and $+$2.5 pp of the changed files.

Since size includes the test changes, these curves describe \acp{PR} as a reviewer sees them: among \acp{PR} of a few dozen lines, agent-supported ones include tests noticeably more often.

\paragraph{The difference depends on the repository, and is smallest where testing is the norm.} The difference varies between repositories for all four outcomes. In every model, $\hat{\tau}$ exceeds $\hat{\beta}$ (Table~\ref{tab:rq1-models}), so the size of the difference depends on the project, and in some projects so does its direction. For $\mathit{IsTestPR}$, the odds ratios of 95\% of repositories are expected to lie between 0.24 and 11.33. Agent-supported \acp{PR} score higher in an estimated 70\% of repositories for $\mathit{IsTestPR}$, 61\% for $\mathit{IsTestOnlyPR}$, 70\% for $R_F$, and 75\% for $R_{\mathit{LOC}}$.

\begin{figure}[t]
    \centering
    \includegraphics[width=\textwidth]{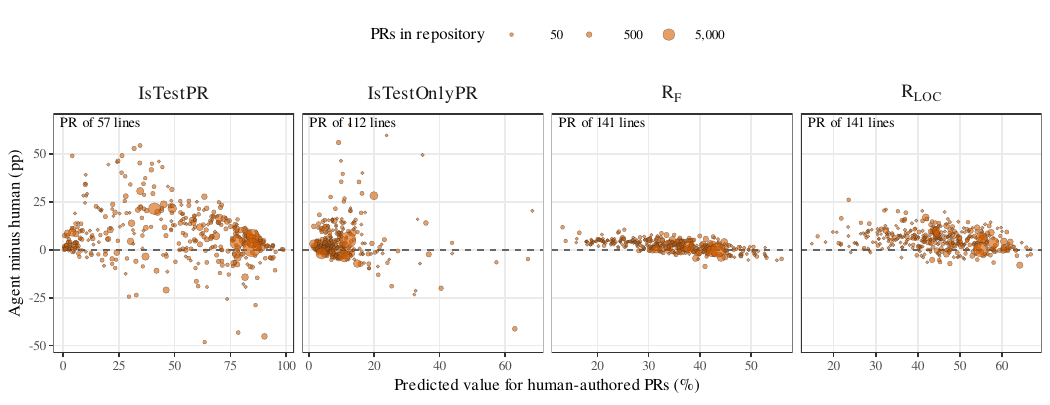}
    \caption{Difference between agent-supported and human-authored \acp{PR} in each repository, against the predicted value for human-authored \acp{PR} in the same repository. Points above the dashed line are repositories where agent-supported \acp{PR} score higher}
    \label{fig:rq1-repo-scatter}
\end{figure}

Part of this variation follows how much a repository already tests. In all four models, the correlation $\hat{\rho}$ between a repository's baseline and its difference is negative ($-$0.28 to $-$0.59): the more a repository tests, the smaller the difference (Figure~\ref{fig:rq1-repo-scatter}). For testing frequency, the median difference is $+$13.8 pp in the third of repositories where human-authored \acp{PR} change test code 33-69\% of the time, against $+$5.6 pp in the lowest third and $+$2.7 pp in the highest. Where testing is the norm, the two groups differ little; where it is rare, the difference is also smaller. This pattern is partly an effect of the percentage scale, on which differences shrink near 0\% and 100\%, but the negative $\hat{\rho}$ shows that it also holds on the log-odds scale. For the extent of testing, where shares are far from these bounds, the pattern is clearer. From the lowest to the highest third, the median difference falls from $+$6.7 to $+$3.5 pp in $R_{\mathit{LOC}}$, and from $+$4.2 to $+$0.3 pp in $R_F$.

\paragraph{Model fit.} All four models converged, and simulation-based residual diagnostics show a close fit for the binary outcomes and a good fit for the shares. The complete data is available in our replication package \citep{milanese2026r}.

\vspace{0.5em}
\noindent\fbox{\parbox{\dimexpr\linewidth-2\fboxsep-2\fboxrule}{%
\textbf{Answer to RQ$_1$:} Agent-supported \acp{PR} include tests more often than human-authored \acp{PR} of the same size: in a typical repository, about 9 more in every 100 \acp{PR}. When they also change production code, more of their changed lines are tests. They are also somewhat more likely to change only tests. The difference is most visible in changes of a few dozen lines, which human developers often submit without tests, and in projects where testing is common but not systematic. It is smaller where testing is already the norm or rare. Comparisons that pool all projects miss these differences, because agent-supported \acp{PR} come mostly from projects that test less.}}

\subsection{RQ$_2$: Type and Context of Testing}
\label{sec:rq-2-results}

Agent-supported and human-authored \acp{PR} differ little in how often they pair test changes with production changes, but they change the test suite differently: within the same repository, agent-supported \acp{PR} add new tests more often, while human-authored \acp{PR} more often modify and delete existing ones. We first show these differences, then how they depend on the context, why pooled comparisons distort them, how they translate into practice for \acp{PR} of different sizes, and how they vary between repositories.

\paragraph{Agent-supported \acp{PR} co-evolve tests slightly more often.} In a typical repository, 57.5\% of mixed agent-supported \acp{PR} change at least one test linked to their production changes, against 55.1\% of human-authored ones of the same size ($+$2.4 pp, $OR$ 1.12, $p_{adj} = 0.034$). The odds ratio is the smallest of the ten effects, and its interval nearly reaches 1, so the two groups differ much less in \emph{whether} they test the production code they change than in \emph{how} they change the tests.

\paragraph{Agent-supported \acp{PR} add tests more often, and modify and delete them less often.} In every context, agent-supported \acp{PR} add tests more often and modify and delete them less often (Table~\ref{tab:rq2-models}). All nine differences remain significant after the Holm-Bonferroni correction. In a typical repository, agent-supported \acp{PR} add a test in about 7 to 10 more of every 100 \acp{PR} ($OR$ 1.50-1.58), modify one in 5 to 8 fewer ($OR$ 0.70-0.76), and delete one in 1 to 4 fewer ($OR$ 0.51-0.77). Modifications include tests that change only through their helper methods. Since the two groups differ little in how often they co-evolve tests, these differences do not stem from co-evolving more: relative to human-authored \acp{PR}, agent-supported \acp{PR} more often extend the test suite with new tests and less often adapt the existing ones.

\begin{table}[t]
    \centering
    \caption{Difference between agent-supported and human-authored \acp{PR} within the same repository, by task and context. The \textit{Human} and \textit{Agent} columns report the predicted share of \acp{PR} in a typical repository.}
    \label{tab:rq2-models}
    \small
    \setlength{\tabcolsep}{3.5pt}
    \begin{tabular}{llrrrrrr}
    \toprule
    Outcome & Context & OR (95\% CI) & $p_{adj}$ & Human & Agent & AME (95\% CI) & $\hat{\tau}$ \\
    \midrule
    \multicolumn{2}{l}{$\mathit{IsCoEvolutionPR}$} & 1.12 (1.01, 1.25) & $0.034$ & 55.1\% & 57.5\% & $+$2.4 (0.2, 4.6) & 0.52 \\
    \addlinespace
    \multirow{3}{*}{$\mathit{Add}$} & $\mathit{TFC}$ & 1.50 (1.20, 1.87) & $0.001$  & 40.3\% & 46.9\% & $+$6.6 (3.0, 10.3)  & \multirow{3}{*}{0.47} \\
                                    & $\mathit{COC}$ & 1.56 (1.42, 1.72) & $<0.001$ & 34.5\% & 43.3\% & $+$8.8 (7.0, 10.7)  & \\
                                    & $\mathit{TFM}$ & 1.58 (1.44, 1.73) & $<0.001$ & 47.3\% & 57.7\% & $+$10.4 (8.3, 12.5) & \\
    \addlinespace
    \multirow{3}{*}{$\mathit{Mod}$} & $\mathit{TFC}$ & 0.70 (0.56, 0.87) & $0.003$  & 72.6\% & 66.1\% & $-$6.5 ($-$10.5, $-$2.5) & \multirow{3}{*}{0.56} \\
                                    & $\mathit{COC}$ & 0.76 (0.69, 0.85) & $<0.001$ & 26.4\% & 21.7\% & $-$4.7 ($-$6.5, $-$2.9)  & \\
                                    & $\mathit{TFM}$ & 0.73 (0.66, 0.80) & $<0.001$ & 52.8\% & 45.0\% & $-$7.9 ($-$10.3, $-$5.4) & \\
    \addlinespace
    \multirow{3}{*}{$\mathit{Del}$} & $\mathit{TFC}$ & 0.51 (0.38, 0.69) & $<0.001$ & 8.1\% & 4.6\% & $-$3.5 ($-$5.0, $-$2.1) & \multirow{3}{*}{0.48} \\
                                    & $\mathit{COC}$ & 0.77 (0.66, 0.91) & $0.003$  & 4.6\% & 3.7\% & $-$0.9 ($-$1.5, $-$0.4) & \\
                                    & $\mathit{TFM}$ & 0.61 (0.52, 0.71) & $<0.001$ & 7.2\% & 4.6\% & $-$2.5 ($-$3.3, $-$1.8) & \\
    \bottomrule
    \end{tabular}
\end{table}

\paragraph{We find no evidence that the difference in additions and modifications depends on the context.} For additions and modifications, we detect no difference between the odds ratios of tests linked to the production changes, of tests that accompany unrelated production changes, and of tests in test-only \acp{PR} ($\chi^2 = 0.21$ and $1.16$, $p_{adj} = 1.000$). The higher likelihood of adding tests is therefore not confined to a particular situation, although this test does not establish that the differences are equal across contexts.

\paragraph{The gap in deletions is smallest for tests linked to the production changes.} Deletions are the only task whose difference depends on the context ($\chi^2 = 10.88$, $p_{adj} = 0.013$). The gap is smallest for tests linked to the production changes: there, the odds ratio is 1.27 times that of unlinked tests in mixed \acp{PR} (95\% CI 1.07-1.51), and the odds ratio of test-only \acp{PR} is 0.67 times that of linked tests (95\% CI 0.49-0.90). A linked deletion is often forced by the production change itself, for example, when the tested method is removed \citep{marsavina2014}. Such deletions are needed whoever authors the \ac{PR}, while deletions that clean up the test suite are not. Although we did not verify that the linked production methods were removed, the pattern is consistent with agent-supported \acp{PR} deleting tests mainly when the change requires it, and less often otherwise.

\begin{figure}[t]
    \centering
    \includegraphics[width=\textwidth]{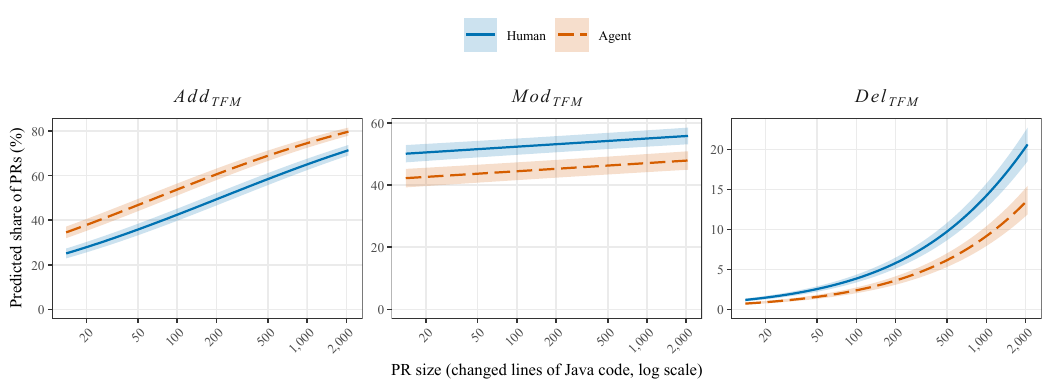}
    \caption{Predicted share of agent-supported and human-authored mixed \acp{PR} performing each task in a typical repository, by size of the \ac{PR}. Rows correspond to tasks and columns to contexts}
    \label{fig:rq2-size-curves}
\end{figure}

\paragraph{In practice, additions and modifications differ at every size, deletions only in larger \acp{PR}.} Figure~\ref{fig:rq2-size-curves} shows the predicted share of mixed \acp{PR} performing each task on unlinked tests in a typical repository across sizes. For additions and modifications, the gap stays between about 4 and 11 pp at most sizes, so a reviewer will see it in small and large \acp{PR} alike. The exception is test-only \acp{PR} that add tests, where the gap narrows for large \acp{PR}, which almost always add tests in both groups (90.2\% of agent-supported and 86.1\% of human-authored \acp{PR} at 500 lines). Deletions instead depend on size. Below 100 lines, mixed \acp{PR} rarely delete tests in either group, and the gap stays under 2 pp. At 1,000 lines, human-authored \acp{PR} delete unlinked tests in 14.2\% of the cases, against 9.2\% for agent-supported ones, while the gap for linked tests is only 2.0 pp. In test-only \acp{PR}, the gap reaches 9.4 pp already at 500 lines (22.2\% vs.\ 12.8\%). Deleting tests is thus mostly part of large changes, and this is where the gap between the two groups appears.

\begin{figure}[t]
    \centering
    \includegraphics[width=\textwidth]{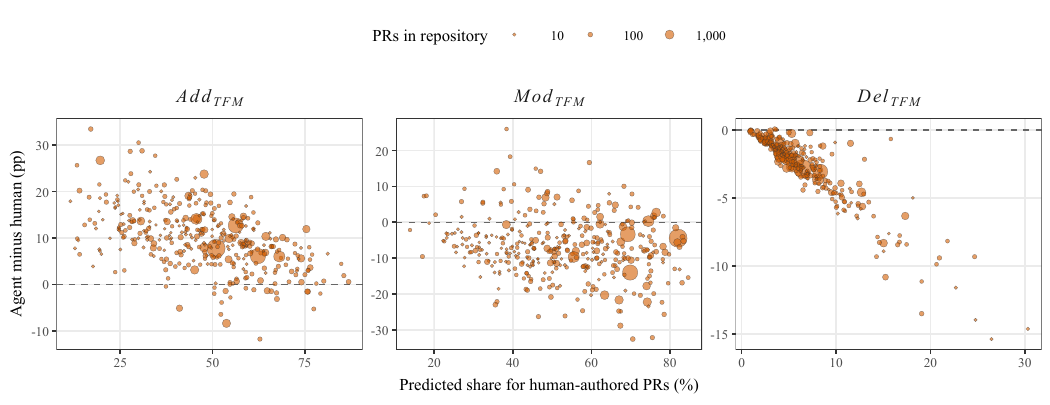}
    \caption{Difference between agent-supported and human-authored mixed \acp{PR} in each repository, against the predicted value for human-authored \acp{PR} in the same repository. Points above the dashed line are repositories where agent-supported \acp{PR} perform the task more often}
    \label{fig:rq2-repo-scatter}
\end{figure}

\paragraph{Agent-supported \acp{PR} add tests more often in most projects, but their maintenance gap follows the project.} As in \hyperref[sec:rq-1-results]{RQ$_1$}, the difference varies between repositories (Figure~\ref{fig:rq2-repo-scatter}), but its direction is more consistent. Agent-supported \acp{PR} add tests more often in an estimated 81-83\% of repositories, depending on the context, and the odds ratios of 95\% of repositories are expected to lie between about 0.6 and 3.9. They modify tests more often in only 26-31\% of repositories, and delete them more often in 8-30\%.

How large the difference is depends on what developers already do in the repository. For additions, the correlation $\hat{\rho}$ between a repository's baseline and its difference is strongly negative ($-$0.91): agents add the most extra tests where developers rarely add tests. In the third of repositories with the highest baseline, the median difference is $+$5.7 to $+$7.3 pp, depending on the context, against $+$8.7 to $+$13.0 pp in the lowest third. For linked tests, the difference peaks in the middle third ($+$10.6 pp). For modifications and deletions, $\hat{\rho}$ is also negative ($-$0.36 and $-$0.45). Since agents perform these tasks less often, their gap widens as the baseline grows. In mixed \acp{PR}, from the lowest to the highest third, the median difference goes from $-$2.7 to $-$7.4 pp for linked modifications, from $-$6.3 to $-$8.0 pp for unlinked ones, and from $-$0.8 to $-$4.0 pp for unlinked deletions. Modifications in test-only \acp{PR} are the exception: two thirds of the repositories modify tests in over 71\% of these \acp{PR}, and the gap narrows towards this ceiling, from $-$8.2 to $-$5.5 pp. In other words, agent-supported \acp{PR} add tests more often whatever the habits of the project, whereas in mixed \acp{PR} their gap in maintenance widens as developers adapt and remove existing tests more often. As in \hyperref[sec:rq-1-results]{RQ$_1$}, part of this pattern comes from the percentage scale, especially for deletions, which are rare, but the negative $\hat{\rho}$ shows that it also holds on the log-odds scale.

\paragraph{Model fit.} All four models converged, and simulation-based residual diagnostics show a close fit. The complete data is available in our replication package \cite{milanese2026r}.

\vspace{1em}
\noindent\fbox{\parbox{\dimexpr\linewidth-2\fboxsep-2\fboxrule}{%
\textbf{Answer to RQ$_2$:} Agent-supported and human-authored \acp{PR} differ little in how often they pair test changes with production changes, but they change the test suite differently. Within the same repository and for \acp{PR} of the same size, agent-supported \acp{PR} add a new test in about 7 to 10 more of every 100 \acp{PR}, and modify or delete an existing one in about 5 to 8 and 1 to 4 fewer. We find no evidence that the differences in additions and modifications depend on whether the tests target the production change, accompany unrelated changes, or make up the whole \ac{PR}. The gap in deletions is smallest for tests linked to the production changes and appears mainly in larger \acp{PR}. Agent-supported \acp{PR} add tests more often in most projects, but in mixed \acp{PR} their gap in maintenance widens where developers adapt and remove existing tests more often.}}

\subsection{RQ$_3$: Test Structure}
\label{sec:rq-3-results}

Within the same repository, the test methods that agent-supported \acp{PR} add differ little in structure from those of human-authored \acp{PR} of the same length, with one exception: they more often have an oracle, in every development context. Some differences appear only in specific contexts, most notably in test-only \acp{PR}, where agents' tests contain more call-assertion sequences. When agent-supported \acp{PR} modify existing test methods, they add at least one assertion more often than human-authored ones, while we detect no difference in how often they remove one. We first report these results, then show how they depend on the context and how they vary between repositories.

\paragraph{Agents' tests differ little from developers' tests in structure.} Table~\ref{tab:rq3-models} reports the difference between the two groups on all outcomes. For eight of the nine outcomes on added test methods, we detect no difference ($p_{adj} = 1.000$). These estimates average over the three development contexts of \hyperref[sec:rq-2-results]{RQ$_2$}, and for three characteristics the difference depends on the context, as we show below. In a typical repository, among test methods of the same length, those added by agent-supported \acp{PR} contain on average 2.39 assertions against 2.36 for human-authored ones, among those with at least one, and 2.24 call-assertion sequences against 2.19. They target 1.35 focal classes against 1.36, and 1.48 focal methods in both groups. They contain conditional logic about as often ($OR$ 0.98), with 1.76 decision points against 1.73 when they do, and create mocks about as often ($OR$ 0.85), with 1.47 mocks in both groups when they do. For these counts, the 95\% intervals of the differences lie within about 6\% of the human mean in either direction (e.g., from $-$1.4\% to $+$3.7\% for assertions), so the data are compatible only with small differences. $\mathit{FocalMethods}$ is the exception, since its interval may be too narrow (see Model fit), so for it we only conclude that we detect no difference. As we did not set an equivalence margin in advance, these bounds describe the precision of the estimates rather than establish equivalence. For whether a test creates mocks, the interval is wide (0.61-1.19). Agent-supported test methods are also slightly shorter (12.1 vs.\ 12.5 lines, ratio of geometric means 0.97, 95\% CI 0.95-0.99).

This contrasts with tests from automated test generators, which are simpler than developer-written tests, consistently target a single focal method, and rely on isolated assertions rather than repeated call-assertion sequences, while the LLM-based generator \textsc{Aster} frequently introduces several mocks \citep{pan2026}. Within a project, the tests that agents add show these traits no more than developers' tests do. Since agents add new tests more often than human developers (\hyperref[sec:rq-2-results]{RQ$_2$}), they extend the test suite with tests structured much like those its developers write.

\begin{table}[t]
    \centering
    \caption{Difference between the test methods of agent-supported and human-authored \acp{PR} within the same repository. The \textit{Human} and \textit{Agent} columns report the predicted value for a typical \ac{PR} in a typical repository. $^{\dagger}$Not adjusted, since $\mathit{Length}$ describes the test methods rather than testing a hypothesis.}
    \label{tab:rq3-models}
    \footnotesize
    \setlength{\tabcolsep}{1.5pt}
    \begin{tabular}{l>{\raggedright\arraybackslash}p{1.5cm}rrrrr}
    \toprule
    Outcome & Ratio (95\% CI) & $p_{adj}$ & Human & Agent & AME (95\% CI) & $\hat{\tau}$ \\
    \midrule
    \multicolumn{7}{l}{\textit{Added test methods}} \\[5pt]
    $\mathit{HasOracle}$      & OR 1.71 (1.29, 2.26)  & $0.001$ & 99.60\% & 99.77\% & $+$0.17 (0.06, 0.27) & 1.17 \\
    $\mathit{Assertions}$     & RR 1.02 (0.98, 1.06)  & $1.000$ & 2.36 & 2.39 & $+$0.03 ($-$0.03, 0.09) & 0.17 \\
    \addlinespace
    $\mathit{HasCondLogic}$   & OR 0.98 (0.85, 1.13)  & $1.000$ & 2.36\% & 2.32\% & $-$0.04 ($-$0.30, 0.22) & 0.17 \\
    $\mathit{DecisionPoints}$ & RR 1.04 (0.96, 1.13)  & $1.000$ & 1.73 & 1.76 & $+$0.03 ($-$0.04, 0.10) & 0.10 \\
    \addlinespace
    $\mathit{Sequences}$      & RR 1.04 (0.99, 1.10)  & $1.000$ & 2.19 & 2.24 & $+$0.05 ($-$0.02, 0.12) & 0.25 \\
    \addlinespace
    $\mathit{CreatesMocks}$ & OR 0.85 (0.61, 1.19) & $1.000$ & 0.94\% & 0.80\% & $-$0.14 ($-$0.43, 0.15) & 1.47 \\
    $\mathit{Mocks}$          & RR 1.00 (0.89, 1.13)  & $1.000$ & 1.47 & 1.47 & 0.00 ($-$0.06, 0.06) & 0.09 \\
    \addlinespace
    $\mathit{FocalClasses}$   & RR 0.97 (0.91, 1.03)  & $1.000$ & 1.36 & 1.35 & $-$0.01 ($-$0.04, 0.01) & 0.28 \\
    $\mathit{FocalMethods}$   & RR 1.00 (0.96, 1.04)  & $1.000$ & 1.48 & 1.48 & 0.00 ($-$0.06, 0.06) & 0.17 \\
    \addlinespace
    $\mathit{Length}$         & GMR 0.97 (0.95, 0.99) & $0.010^{\dagger}$ & 12.5 & 12.1 & $-$0.4 ($-$0.7, $-$0.1) & 0.10 \\
    \midrule
    \multicolumn{7}{l}{\textit{Modified test methods}} \\[5pt]
    $\mathit{RemovesAssertions}$ & OR 1.01 (0.85, 1.20) & $0.941$ & 13.9\% & 14.0\% & $+$0.1 ($-$1.8, 1.9) & 0.70 \\
    $\mathit{AddsAssertions}$    & OR 1.31 (1.09, 1.57) & $0.007$ & 19.6\% & 23.5\% & $+$3.8 (1.2, 6.5) & 0.73 \\
    $\mathit{LosesOracle}$ & OR 0.008 (0.001, 0.056) & $<0.001$ & 0.16\% & $<$0.01\% & $-$0.16 ($-$0.22, $-$0.09) & 6.29 \\
    \bottomrule
    \end{tabular}
\end{table}

\paragraph{Agents' tests more often have an oracle.} Among the nine tested outcomes, whether a test method has an oracle is the only structural characteristic on which the two groups differ significantly ($OR$ 1.71, 95\% CI 1.29-2.26, $p_{adj} = 0.001$). Since most test methods in both groups have an oracle, and \acp{PR} differ strongly in how many of their tests lack one, the predicted share for a typical \ac{PR} is close to 100\% in both groups (99.77\% vs.\ 99.60\%). The difference lies in the \acp{PR} whose tests often lack an oracle: in our sample, 6.4\% of the test methods added by agent-supported \acp{PR} have no oracle, against 10.6\% of those of human-authored ones. A test without an oracle fails only if the code it exercises throws an exception, so it adds to the coverage of the code without checking its behavior \citep{inozemtseva2014}. Such tests are concentrated in few \acp{PR}: repositories and \acp{PR} together account for 84\% of the variation in whether a test method has an oracle, so a \ac{PR} tends to add either tests with oracles or tests without them. Since we only count the assertions in a test method's own body, a test method that delegates its checks to helper methods has no explicit oracle by our definition. Human-authored test methods call helpers somewhat more often (24.8\% vs.\ 20.6\%), so part of this difference may reflect how the two groups use helpers rather than whether their tests check behavior.

\paragraph{Agents' tests have an oracle more often in every context, but other differences depend on the context.} For five of the nine characteristics, the difference between the two groups is the same whether the tests are linked to the production changes ($\mathit{COC}$), accompany unrelated production changes ($\mathit{TFM}$), or make up a test-only \ac{PR} ($\mathit{TFC}$) ($\chi^2$ from 3.77 to 6.47, $p_{adj} \geq 0.197$). In particular, agents' tests have an oracle more often in all three contexts ($OR$ 1.82, 1.61, and 2.42, respectively). For three others, the difference depends on the context. Call-assertion sequences ($\chi^2 = 36.77$, $p_{adj} < 0.001$) differ mainly in test-only \acp{PR}, where agents' tests contain more of them ($RR$ 1.32, 95\% CI 1.15-1.52), while in mixed \acp{PR} the two groups are close ($RR$ 0.98 and 1.06). Conditional logic ($\chi^2 = 38.51$, $p_{adj} < 0.001$) differs in opposite directions within mixed \acp{PR}: agents' tests contain it less often when they are linked to the production changes ($OR$ 0.80, 95\% CI 0.68-0.94) and more often when they are not ($OR$ 1.18, 95\% CI 1.01-1.38), which offsets the two in the overall estimate. For focal methods ($\chi^2 = 23.81$, $p_{adj} < 0.001$), the ratios in all contexts remain close to 1 (0.97 to 1.04), so the dependence is detectable but small. Whether a test creates mocks also appears to depend on the context ($\chi^2 = 46.68$), with agents' tests creating mocks more often in test-only \acp{PR} ($OR$ 2.14). Since this model could not account for the variation between \acp{PR} (see Model fit), its test and intervals are likely too narrow, and we do not interpret this result. In test-only \acp{PR}, where a \ac{PR}'s only purpose is to change the test suite, the tests that agents add thus consist of more call-assertion blocks than those of human developers, while targeting as many focal methods ($RR$ 1.00, 95\% CI 0.89-1.13). If anything, this moves them further from the isolated assertions typical of automated generators.

\paragraph{When modifying tests, agents add assertions more often.} Although agent-supported \acp{PR} modify existing tests less often than human-authored ones (\hyperref[sec:rq-2-results]{RQ$_2$}), when they do, they add at least one assertion more often: in a typical repository, to 23.5\% of the test methods they modify, against 19.6\% for human-authored \acp{PR} with edits of the same size ($OR$ 1.31, 95\% CI 1.09-1.57, $p_{adj} = 0.007$). Since an assertion whose code changes counts as one removed and one added assertion, these additions include both new and rewritten assertions, which our outcomes do not distinguish. We detect no difference in how often agent-supported \acp{PR} remove or rewrite at least one assertion ($OR$ 1.01, 95\% CI 0.85-1.20).

Coding agents have been observed to make failing tests pass by editing their expected values or deleting them rather than fixing the code \citep{anthropic2025claude37, zhong2026}. Our outcomes do not measure whether a change weakens what a test checks, since an added assertion may be redundant and a rewritten one may be stronger. They do capture, however, the most direct syntactic form of weakening: a test method that loses its explicit oracle altogether. This is rare in both groups, affecting 0.79\% of the test methods modified by agent-supported \acp{PR} (458) and 0.49\% of those modified by human-authored ones (1,658), and it is strongly concentrated. Most \acp{PR} lose no oracle and a few lose many, so the variation between \acp{PR} cannot be estimated and we fit the model without the \ac{PR} term (see Model fit), and the difference between the two groups varies between repositories far more than for any other outcome ($\hat{\tau} = 6.29$). For a typical repository, the model estimates that agent-supported \acp{PR} lose the oracle of a modified test much less often ($OR$ 0.008, 95\% CI 0.001-0.056, $p_{adj} < 0.001$). The repository-specific estimates, however, show a different picture: in most repositories they lie close to this value, whereas agents lose the oracle more often in an estimated 22\% of repositories, and in seven repositories the repository-specific interval lies entirely above 1, against none entirely below. Since a single odds ratio does not summarize this pattern, we do not interpret it as evidence that agents lose oracles less often. Rather, losing the oracle is rare and concentrated in a few projects, in some of which agent-supported \acp{PR} do so far more often than human-authored ones. Since we only count a test method's own assertions, these losses may also include refactorings that move its checks into helper methods, which we did not inspect. Our data therefore show neither that agents weaken the tests they modify more often than developers, nor that they do not.

\paragraph{Test structure varies more between projects than between the two groups.} For every outcome, the baseline varies more between repositories than the difference between the two groups does: the standard deviation of the repository baselines exceeds $\hat{\tau}$ in all models. For the counts, $\hat{\tau}$ is small (0.09 to 0.28): for example, the rate ratios of 95\% of repositories are expected to lie between 0.73 and 1.41 for assertions, and between 0.72 and 1.39 for focal methods. This is consistent with agents following the conventions of each project, although our data cannot distinguish this from other explanations, such as the developers who steer the agents or the reviewers who accept their \acp{PR} enforcing these conventions. Whether a test has an oracle varies more ($\hat{\tau} = 1.17$, 95\% of repository odds ratios between 0.17 and 16.9), but its direction is consistent: agents' tests have an oracle more often in an estimated 68\% of repositories, and no repository has an interval entirely below 1, against 12 with an interval entirely above it. As in \hyperref[sec:rq-1-results]{RQ$_1$}, the difference is smallest where tests already have an oracle ($\hat{\rho} = -0.56$). For modified tests, agents add assertions more often in an estimated 64\% of repositories, and remove them more often in 50\%.

\paragraph{Model fit.} All models converged, and residual diagnostics show a close fit for the binary outcomes and a good fit for the counts, except that the dispersion of $\mathit{FocalMethods}$ is slightly underestimated, so its interval may be too narrow, and that of $\mathit{Sequences}$ overestimated, so its bound is conservative. $\mathit{CreatesMocks}$ and $\mathit{LosesOracle}$ were fitted without the \ac{PR} term, since the variation between \acp{PR} could not be estimated, so their intervals may be too narrow; the complete diagnostics are available in our replication package \citep{milanese2026r}.

\vspace{0.5em}
\noindent\fbox{\parbox{\dimexpr\linewidth-2\fboxsep-2\fboxrule}{%
\textbf{Answer to RQ$_3$:} Within the same project and at equal length, the tests that agent-supported \acp{PR} add differ little in structure from those that developers write. The one consistent difference is that agents' tests more often have an oracle, in every context. Other differences depend on the context. When agents modify existing tests, they more often add at least one assertion, and we detect no difference in how often they remove one. Losing the oracle altogether is rare in both groups and concentrated in a few projects, in some of which agents do so far more often than developers. Test structure varies more between projects than between agents and developers, which is consistent with agents following each project's conventions.}}

\subsection{RQ$_4$: Test Smells}
\label{sec:rq-4-results}

Agent-supported and human-authored \acp{PR} show different patterns of test-smell evolution depending on the testing activity, although the evidence is weaker than for the previous research questions. At the PR level, agent-supported additions exhibit a smaller overall Smell Delta, whereas modifications exhibit a larger one. These differences, however, are close to the significance threshold and are not reproduced at the level of individual test methods. We first examine the aggregate \ac{PR}-level differences, then the individual smell categories, and finally whether the same patterns persist at the method level.

\paragraph{Agent-supported additions and modifications show opposite \ac{PR}-level patterns.} Table~\ref{tab:rq4-pr-smells} reports the \ac{PR}-level results. For test modifications, agent-supported \acp{PR} exhibit a higher overall Smell Delta than human-authored ones ($\beta=10.829$, $p=0.046$). After accounting for change size and repository-level variability, agent-supported modifications are therefore associated with a larger net increase in test-smell occurrences.

For test additions, the direction is reversed. Agent-supported \acp{PR} exhibit a lower overall Smell Delta than human-authored ones ($\beta=-8.271$, $p=0.024$), indicating a smaller net increase in smell occurrences. Both aggregate estimates are close to the significance threshold, so we interpret their magnitude and direction together with the category-specific evidence rather than relying on statistical significance alone.

\begin{table}[htb!]
    \centering
    \scriptsize
    \caption{PR-level differences in test-smell evolution between
    agent-supported and human-authored \acp{PR}. Positive $\beta$ values
    indicate a larger Smell Delta for agent-supported contributions, whereas
    negative values indicate a smaller one. Category-specific $p$-values are
    Holm-adjusted; the overall Smell Delta is analyzed separately.}
    \label{tab:rq4-pr-smells}
    \begin{tabular}{llrl}
        \toprule
        Change & Test Smell &
        $\beta$ (95\% CI) &
        $p$ \\
        \midrule
        Modification & Overall Smell Delta
        & 10.829 (0.204, 21.453)
        & 0.046 \\

        \midrule
        Addition & Overall Smell Delta
        & $-$8.271 ($-$15.440, $-$1.102)
        & 0.024 \\

        \midrule
        Addition & General Fixture
        & $-$0.598 ($-$0.915, $-$0.282)
        & 0.004 \\

        Addition & Exception Catching Throwing
        & $-$0.458 ($-$0.718, $-$0.197)
        & 0.011 \\
        \bottomrule
    \end{tabular}
\end{table}

\paragraph{Only two smell categories differ after correction.} The category-specific results provide more limited evidence of systematic differences. For additions, most coefficients are negative, but only \emph{General Fixture} ($\beta=-0.598$, $p_{adj}=0.004$) and \emph{Exception Catching Throwing} ($\beta=-0.458$, $p_{adj}=0.011$) remain significant after Holm correction. Other categories, including \emph{Magic Number Test}, \emph{Lazy Test}, and \emph{Unknown Test}, also have negative coefficients but do not remain significant after correction.

\begin{figure*}[t]
    \centering
    \includegraphics[width=\textwidth]
        {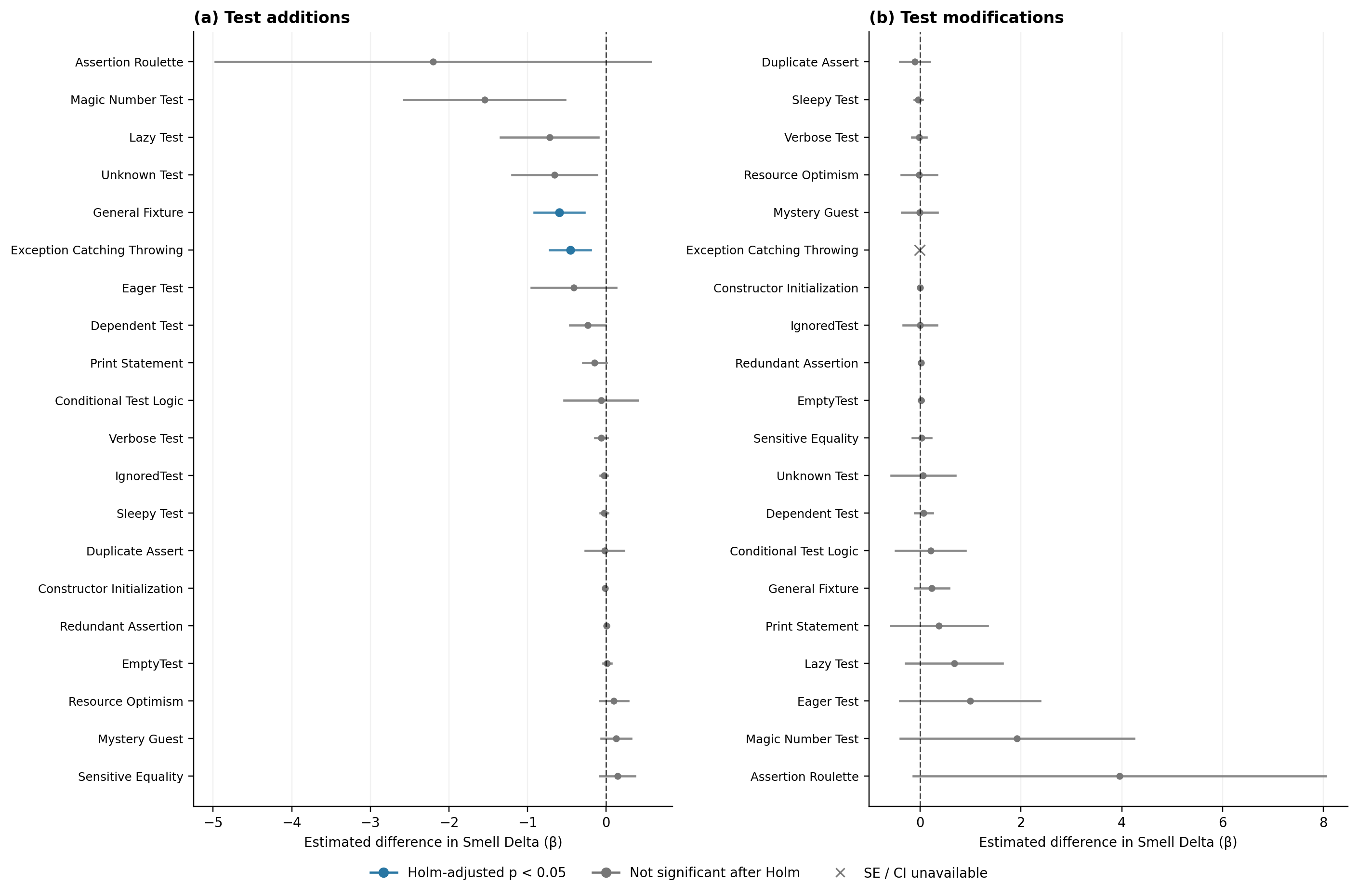}
    \caption{Estimated \ac{PR}-level differences in test-smell evolution between
    agent-supported and human-authored \acp{PR} for (a) test additions and
    (b) test modifications. Points indicate regression coefficients
    ($\beta$), with 95\% Wald \acp{CI}. Blue points identify categories
    significant after Holm correction ($p_{adj}<0.05$), whereas
    gray points indicate non-significant differences. Crosses identify
    estimates for which standard errors and \acp{CI} were unavailable.}
    \label{fig:rq4-pr-categories}
\end{figure*}

For modifications, no individual smell category differs significantly after correction. Several estimates are positive, including \emph{Assertion Roulette} ($\beta=3.960$) and \emph{Magic Number Test} ($\beta=1.924$), indicating that the positive aggregate difference is not attributable to a single smell category. Figure~\ref{fig:rq4-pr-categories} shows the complete set of category-specific estimates.

\paragraph{The \ac{PR}-level patterns do not persist at the method level.} When individual test methods are considered, we detect no statistically significant aggregate difference between agent-supported and human-authored contributions. For modifications, the estimated coefficient is negative ($\beta=-0.0175$, $p=0.101$), whereas for additions it is positive ($\beta=0.0535$, $p=0.229$).

The category-specific results lead to the same conclusion. For modifications, \emph{Duplicate Assert} ($\beta=-0.00156$, $p=0.016$) and \emph{Verbose Test} ($\beta=-0.00073$, $p=0.043$) are nominally significant, but neither remains so after Holm correction. For additions, no individual smell category is statistically significant even before correction.

\begin{table}[htb!]
    \centering
    \scriptsize
    \caption{Aggregate method-level differences in test-smell evolution
    between agent-supported and human-authored \acp{PR}.}
    \label{tab:rq4-method-smells}
    \begin{tabular}{lrl}
        \toprule
        Change &
        $\beta$ (95\% CI) &
        $p$ \\
        \midrule
        Modification
        & $-$0.0175 ($-$0.0384, 0.0034)
        & 0.101 \\

        Addition
        & 0.0535 ($-$0.0337, 0.1408)
        & 0.229 \\
        \bottomrule
    \end{tabular}
\end{table}

\begin{figure*}[t]
    \centering
    \includegraphics[width=\textwidth]
        {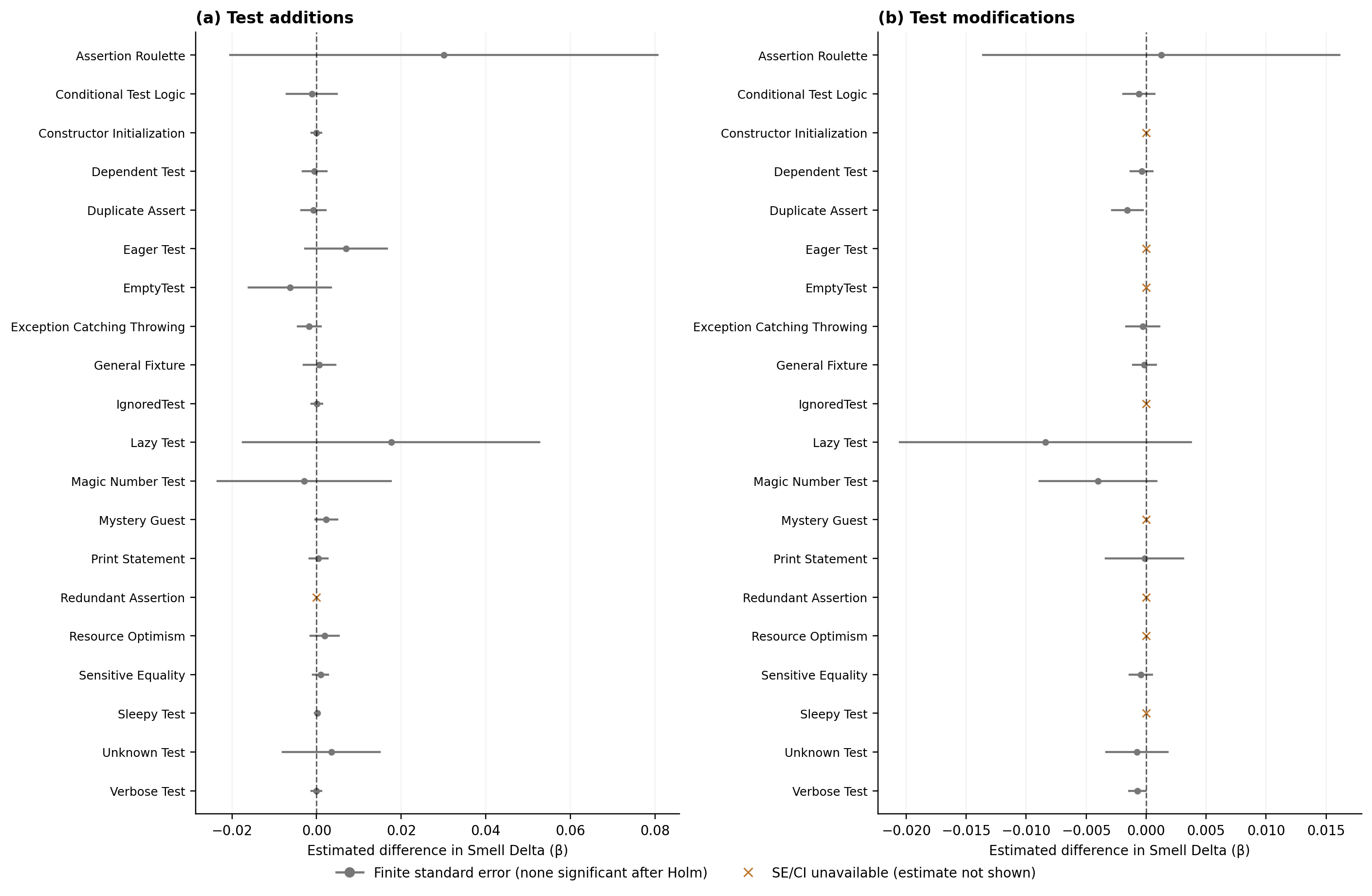}
    \caption{Estimated method-level differences in test-smell evolution
    between agent-supported and human-authored \acp{PR} for (a) test
    additions and (b) test modifications. Points and error bars represent
    regression coefficients and 95\% Wald \acp{CI}. None of the
    category-specific differences remains significant after Holm correction.
    Crosses mark categories with unavailable standard errors.}
    \label{fig:rq4-method-categories}
\end{figure*}

Figure~\ref{fig:rq4-method-categories} reports the corresponding category-specific estimates.

\paragraph{The two levels disagree.} At the PR level, modifications have a positive coefficient and additions a negative one, whereas the corresponding method-level coefficients have the opposite signs and are not significant. The two levels differ in which smell occurrences they attribute to a contribution and in how they weight \acp{PR}; we discuss possible explanations in Section~\ref{sec:disc-quality}.

\paragraph{Model estimation.} The aggregate models yield estimates and standard errors at both levels. Some category-specific method-level models do not, and we do not interpret them.

\vspace{1em}
\noindent\fbox{\parbox{\dimexpr\linewidth-2\fboxsep-2\fboxrule}{%
\textbf{Answer to RQ$_4$:} Agent-supported and human-authored contributions show different patterns of test-smell evolution at the PR level: agent-supported additions exhibit a smaller overall Smell Delta, whereas modifications exhibit a larger one, although both differences are close to the significance threshold. Only \emph{General Fixture} and \emph{Exception Catching Throwing}, both in additions, differ significantly after correction. These differences are not reproduced at the level of individual test methods, where neither the aggregate nor the category-specific analyses provide evidence of systematic differences.}}

\subsection{RQ$_5$: Code Coverage}
\label{sec:rq-5-results}

Among the test \acp{PR} we could execute, we detect no difference in how well agent-supported and human-authored \acp{PR} cover their production changes. The only significant difference is in test-only \acp{PR}, where the tests that agents contribute cover a smaller part of the code they exercise. We first report the coverage of production changes and of the contributed tests, and then how the coverage of the whole codebase changes. Finally, we show how these results depend on the code that the \acp{PR} start from, on their size, and on the repository, and how robust they are.

\paragraph{We detect no difference in the coverage of production changes.} Table~\ref{tab:rq5-models} reports the difference between the two groups on all metrics. In a typical repository, when a test \ac{PR} also changes production code, the test suite covers 50.7\% of the lines of the changed classes in agent-supported \acp{PR} and 49.5\% in human-authored ones that start from the same coverage ($+$1.2 pp, $OR$ 1.07, $p_{adj} = 1.000$). Similarly, it executes 51.3\% of the production lines that agent-supported \acp{PR} add, against 48.1\% for human-authored ones ($+$3.2 pp, $OR$ 1.14, $p_{adj} = 1.000$). Neither difference is significant, but this does not make the two groups equivalent. For new lines, the data are compatible with differences ranging from a disadvantage for agents of about 3 pp to an advantage of about 9 pp. For the changed classes, the interval is narrower ($-$1.2 to $+$3.6 pp), but it depends on a dispersion structure that our diagnostics question, so we do not rely on its width. In both groups, the tests contributed in the \ac{PR} reach about three quarters of the new lines that are covered (77.4\% vs.\ 75.4\%, $OR$ 1.12, $p = 0.343$). Hence, in both groups most covered new lines are executed by the tests that accompany the change, although pre-existing tests may execute them as well. Therefore, although agents more often add new tests and less often adapt existing ones (\hyperref[sec:rq-2-results]{RQ$_2$}), we find no sign that this leaves the production changes of their test \acp{PR} less covered.

These results only concern production changes that come with test changes. Production-only \acp{PR} are not part of this analysis, and agent-supported \acp{PR} change test code more often than human-authored ones (\hyperref[sec:rq-1-results]{RQ$_1$}). The two groups of test \acp{PR} may therefore differ in which production changes they include, and our results do not show whether agent-supported \acp{PR} leave production code better or worse covered overall.

\begin{table}[t]
    \centering
    \caption{Difference in coverage between agent-supported and human-authored test \acp{PR} within the same repository. The \textit{Human} and \textit{Agent} columns report the predicted coverage in a typical repository.}
    \label{tab:rq5-models}
    \footnotesize
    \setlength{\tabcolsep}{2.2pt}
    \begin{tabular}{>{\raggedright\arraybackslash}p{2.1cm}rrrrrrr}
    \toprule
    Metric & Type & OR (95\% CI) & $p_{adj}$ & Human & Agent & AME (95\% CI) & $\hat{\tau}$ \\
    \midrule
    \multirow{2}{*}{\textit{Affected-Class}}  & TFC       & 0.75 (0.51, 1.09) & $0.651$ & 78.4\% & 73.8\% & $-$4.7 ($-$10.6, 1.5)     & \multirow{2}{*}{0.64} \\
                                           & Mixed     & 0.96 (0.79, 1.16) & $1.000$ & 62.1\% & 61.3\% & $-$0.8 ($-$4.1, 2.4)      & \\
    \addlinespace
    \multirow{2}{*}{\textit{Contributed-Test}}      & TFC       & 0.62 (0.44, 0.87) & $0.032$ & 55.6\% & 44.5\% & $-$11.1 ($-$18.1, $-$3.0) & \multirow{2}{*}{0.24} \\
                                           & Mixed     & 0.95 (0.83, 1.10) & $1.000$ & 53.1\% & 51.9\% & $-$1.1 ($-$4.4, 2.4)      & \\
    \addlinespace
    \textit{Production-Change}                & Mixed & 1.07 (0.93, 1.22) & $1.000$ & 49.5\% & 50.7\% & $+$1.2 ($-$1.2, 3.6)      & 0.38 \\[15pt]
    \multicolumn{3}{l}{\textit{New-Line Coverage}} \\
    $|N_A|\,/\,|N|$                        & Mixed & 1.14 (0.90, 1.45) & $1.000$ & 48.1\% & 51.3\% & $+$3.2 ($-$2.6, 8.9)      & 0.73 \\
    $|N_T|\,/\,|N_A|$                      & Mixed     & 1.12 (0.89, 1.41) & $0.343$ & 75.4\% & 77.4\% & $+$2.0 ($-$2.1, 6.1) & 0.29 \\
    \bottomrule
    \end{tabular}
\end{table}

\paragraph{In test-only \acp{PR}, agents' tests cover less of the code they exercise.} Contributed-Test Coverage is the only metric on which the two groups differ significantly. In test-only \acp{PR} of a typical repository, the tests that agent-supported \acp{PR} contribute cover 44.5\% of the lines of the classes they execute, against 55.6\% for human-authored ones ($-$11.1 pp, $OR$ 0.62, $p_{adj} = 0.032$). In mixed \acp{PR}, we detect no difference (51.9\% vs.\ 53.1\%, $OR$ 0.95), and the difference between the two types of \ac{PR} is itself significant (ratio of odds ratios 1.55, $\chi^2 = 6.24$, $p_{adj} = 0.025$). The lower coverage is not explained by the amount of test code. In this sample, agent-supported and human-authored test-only \acp{PR} change a similar amount of test code (median 58 vs.\ 61 lines), and their tests execute classes of similar size (median 764 vs.\ 662 coverable lines), but agents' tests cover fewer of these lines (median 173 vs.\ 389). This suggests that, in \acp{PR} that only change tests, agents write tests that reach a class but exercise a narrower part of it. Since coverage measures how much code the tests execute rather than whether they detect faults \citep{inozemtseva2014}, a narrower test is not necessarily a weaker one. In test-only \acp{PR}, however, RQ$_3$ detects no difference in the number of focal methods of the tests that agents add ($RR$ 1.00), so the narrower coverage does not reflect a narrower scope as measured by \textsc{Hamster}.

We cannot tell whether the rest of the test suite makes up for this. After the \ac{PR}, we detect no significant difference in how well the whole suite covers the classes that the contributed tests exercise (Affected-Class Coverage, 73.8\% vs.\ 78.4\%, $OR$ 0.75, $p_{adj} = 0.651$), but the interval ranges from a gap of about 11 pp to an advantage of about 2 pp.

The difference in Contributed-Test Coverage is also the least precise of our findings. It rests on 142 \acp{PR} in 45 repositories, but only 11 of these repositories contain test-only \acp{PR} of both groups. The model also draws on the other 34, which inform the repository baselines without providing a comparison within the repository, and it shares $\hat{\tau}$ with mixed \acp{PR}. Within the 11 repositories with both groups (28 agent-supported and 35 human-authored \acp{PR}), the raw difference points the same way: the contributed tests cover on average 30.4\% of their classes in agent-supported \acp{PR} and 43.5\% in human-authored ones, and less in agent-supported \acp{PR} in 7 of the 11 repositories. These counts are too small to assess the influence of individual projects, so this result should be confirmed on a larger sample.

\paragraph{Codebase coverage, including test code, increases more often in agent-supported \acp{PR}.} Unlike the other metrics, codebase coverage covers all the classes of the project, which in our coverage data include test classes. We count it as unchanged when it varies by less than 0.01 pp. In mixed \acp{PR}, it increases in 59.7\% of agent-supported \acp{PR} and 49.8\% of human-authored ones ($OR$ 1.34, 95\% CI 1.02-1.76, $p = 0.037$), and decreases in 26.2\% and 30.0\% ($OR$ 0.74, 95\% CI 0.57-0.95, $p = 0.018$). In test-only \acp{PR}, agent-supported \acp{PR} also raise it more often (53.4\% vs.\ 39.2\%), but neither difference is significant ($OR$ 1.60 and 0.92). Moreover, 40.5\% of human-authored test-only \acp{PR} leave it unchanged, against 20.7\% of agent-supported ones, consistent with human-authored \acp{PR} modifying existing tests more often and adding new ones less often (\hyperref[sec:rq-2-results]{RQ$_2$}). Since test classes are included, a change in this metric also depends on how well the added or removed test code is itself covered compared with the rest of the codebase, not only on how well production code is tested. We therefore report these differences as context only, and do not read them as evidence that agents leave production code better tested.

\paragraph{Agents' tests in mixed \acp{PR} reach code that is already better covered.} To check whether the two groups start from different coverage, we fit a linear mixed model of the pre-merge coverage (on the logit scale) with repository random effects. Within the same repository, the classes affected by mixed agent-supported \acp{PR} start out better covered than those of human-authored ones ($+$0.49, 95\% CI 0.14-0.85, $p = 0.007$), while the changed production classes start from the same coverage ($+$0.02, $p = 0.925$). The difference therefore comes from the other classes that the contributed tests execute: the classes that agents' tests reach tend to be better covered by the existing suite before the \ac{PR}. When two groups start from different values, adjusting for the baseline and analyzing the change can lead to different conclusions \citep{lord1967}. Here, neither approach detects a difference between the two groups ($OR$ 0.96 adjusted for the baseline, and 1.05 on the change in coverage).

\paragraph{In practice, the gap in test-only \acp{PR} is widest in small changes.} In the main models, size has the same effect in both groups: the tests contributed in larger test-only \acp{PR} cover more of the classes they exercise (doubling the total Java churn multiplies the odds by 1.22). To see how the difference between the two groups translates into percentages for \acp{PR} of different sizes, we add an interaction between \texttt{agentic} and \texttt{size} to each model, which we use only for this purpose. In test-only \acp{PR}, the gap is widest for small \acp{PR}: at 10 lines, the tests in agent-supported \acp{PR} cover 30.8\% of their classes, against 46.9\% in human-authored ones ($-$16.1 pp, 95\% CI $-$25.9 to $-$6.4), and at 50 lines 44.3\% against 56.6\% ($-$12.4 pp). From about 200 lines, the difference is no longer significant ($-$8.0 pp at 200 lines and $-$3.0 pp at 1,000 lines). However, the interaction itself is not significant ($p = 0.123$), so a constant odds ratio across sizes cannot be excluded. In mixed \acp{PR}, larger \acp{PR} leave a larger share of their new lines uncovered in both groups (from about 61\% covered at 10 lines to 42\% at 1,000 lines for human-authored \acp{PR}). In small mixed \acp{PR}, the tests contributed by agents also reach a larger share of the covered new lines ($+$12.6 pp at 10 lines and $+$6.9 pp at 50 lines), a difference that vanishes from about 200 lines. We report this interaction ($p = 0.030$) descriptively, since it is not part of any tested family.

\paragraph{The difference depends on the repository, and has no consistent direction.} Unlike in \hyperref[sec:rq-1-results]{RQ$_1$} and \hyperref[sec:rq-2-results]{RQ$_2$}, where agents differed in the same direction in most repositories, the coverage of production changes varies between repositories without a consistent direction. Agent-supported \acp{PR} cover their changed classes better in an estimated 57\% of repositories, with the odds ratios of 95\% of repositories expected to lie between 0.50 and 2.27. Covered new lines vary the most ($\hat{\tau} = 0.73$, 95\% of repository odds ratios between 0.27 and 4.79, agents higher in an estimated 57\%), so whether agents leave more or less of their new code uncovered depends on the project. For Contributed-Test Coverage in test-only \acp{PR}, the few repositories with both groups do not allow us to estimate this variation separately, as discussed above.

\paragraph{Robustness.} We repeat the analyses with nine variations: removing test classes from the coverage data, which is possible for Affected-Class, Contributed-Test, and Production-Change Coverage; keeping all \acp{PR} regardless of how many contributed tests we could verify, or only those where all of them are verified; excluding \acp{PR} in which \textsc{Hamster} did not detect some of the changed test methods; excluding \acp{PR} where a new test fails; modeling the change in coverage instead of adjusting for the baseline; letting the effect of \texttt{agentic} differ for \acp{PR} without a baseline; adding the amount of test code changed; and, for Contributed-Test Coverage, also adding whether the \ac{PR} adds and modifies tests. No conclusion changes in any of them. In particular, the difference in Contributed-Test Coverage in test-only \acp{PR} holds in all seven variations that apply to it ($OR$ 0.55-0.64, all $p < 0.05$). Since the classes that the contributed tests execute include test classes, whose lines are executed by their own tests, the variation without test classes is especially relevant: it leaves the odds ratio unchanged ($OR$ 0.62), so the difference is not caused by tests covering their own classes. The difference also holds when we keep all \acp{PR} regardless of verification ($OR$ 0.55), which matters because the verification excludes more agent-supported than human-authored test-only \acp{PR} (29.9\% vs.\ 16.2\%). Adding the amount of test code changed to the model of covered new lines reduces the small, non-significant advantage of agents from $OR$ 1.14 to 0.99, so any difference there runs through the amount of test code, which is larger in agent-supported \acp{PR} (\hyperref[sec:rq-1-results]{RQ$_1$}).

\paragraph{Model fit.} All models converged with the full random-effects structure, and residual diagnostics show an acceptable fit, except that for Affected-Class and Production-Change Coverage the dispersion varies between \acp{PR} more than our models allow, so we do not rely on the width of their intervals; the complete diagnostics are available in our replication package \citep{milanese2026r}.

\vspace{0.5em}
\noindent\fbox{\parbox{\dimexpr\linewidth-2\fboxsep-2\fboxrule}{%
\textbf{Answer to RQ$_5$:} Among \acp{PR} that change test code, we detect no difference in how well agent-supported and human-authored \acp{PR} cover their production changes, and in both groups about three quarters of the covered new lines are reached by the tests contributed in the \ac{PR}. Agents' tendency to add new tests rather than adapt existing ones therefore does not appear to leave their changes less covered, although our data cannot establish that the two groups are equivalent, nor speak to production changes made without tests. The only significant difference appears in test-only \acp{PR}, where the tests that agents contribute cover about 11 pp less of the classes they exercise, especially in small \acp{PR}. Since only 11 repositories contribute test-only \acp{PR} of both groups, this result should be confirmed on a larger sample. Codebase coverage, which includes test code, increases more often in agent-supported \acp{PR}, but this mostly reflects the test code they add.}}

\section{Discussion}
\label{sec:discussion}

Section~\ref{sec:results} already interprets the results of each research question. This section discusses what they mean together, relates them to prior work, and derives implications for research and practice; it closes with the threats to validity. Since agent-supported \acp{PR} are, by construction, collaborations between developers and agents, the differences we observe characterize this collaboration as a whole. They may reflect how agents carry out a task, but also how developers choose, specify, and review the tasks they delegate, which our data do not separate; we use phrases such as ``agents add tests more often'' in this descriptive sense.

\subsection{Comparisons must account for the project}
\label{sec:disc-within-repo}
\label{sec:disc-heterogeneity}

Pooled and within-repository comparisons of the same contributions lead to different conclusions (RQ$_1$). Agent-supported \acp{PR} are concentrated in repositories whose \acp{PR} change test code less often, so a pooled comparison mixes two questions: whether agent-supported \acp{PR} differ from human-authored ones within a project, and in which projects coding agents are adopted. Our previous study pooled \acp{PR} across repositories~\citep{milanese2026}, and the other comparisons of agentic and human tests pool them at least in part (Section~\ref{sec3}). Our results do not invalidate their findings, which concern different populations and outcomes, but they show that project composition can be large enough to hide a substantial difference and to reverse the sign of others.

The project matters in a second way. The average differences we report summarize a distribution of repository-specific differences whose spread is often larger than the average itself, so statements such as ``agent-supported \acp{PR} test more'' hold on average, not necessarily in a given project. Part of this variation is systematic: for testing frequency, test additions, and oracles, the difference is smallest where the practice is already common in the repository, which is consistent with agents, or the developers who steer them, following each project's conventions. The rest is unexplained, and candidate moderators include the maturity of the test suite, the strictness of continuous integration and review, the presence of agent instruction files, and the agent used.

\subsection{Agents take on a different share of testing work}
\label{sec:disc-accretive}

Taken together, RQ$_2$, RQ$_3$, and RQ$_5$ suggest that the two groups differ more in which testing tasks they carry out than in the tests they write. Agent-supported \acp{PR} add tests more often and modify or delete them less often, but we detect no difference in most structural characteristics of the tests they add, nor in how well their production changes are covered. These are differences in whether a \ac{PR} performs a task, not in how many tests it adds or how fast a suite grows.

A possible explanation is that adding a test is a self-contained task that can be carried out from the context of the change at hand, whereas modifying or deleting an existing test requires reconstructing what it checks and judging whether it is still needed. Two further explanations are plausible but untested: agents may be instructed to add tests for the changes they make rather than to revise existing ones, and developers may delegate to agents tasks that call for new tests. Whatever the cause, if agent adoption grows, the maintenance of existing tests may shift toward developers, and obsolete tests may accumulate where agents contribute most; whether this happens can only be established by a longitudinal analysis of test changes over the history of each project.

Our coverage results are not directly comparable with those of prior studies. \citet{alves2026test} report higher diff coverage for AI-only than for human \acp{PR} in Python, comparing \acp{PR} pooled across repositories, whereas within repositories we detect no difference in the coverage of added lines. \citet{dipongkor2026coverage} run the post-PR suite with and without the test changes of a \ac{PR}, so that production code is fixed and only the tests vary, and estimate the coverage gained through those changes. We instead compare two complete revisions and attribute coverage to individual test methods, so our share of covered added lines executed by contributed tests also includes lines that pre-existing tests execute.

\subsection{Test quality remains an open question}
\label{sec:disc-quality}
\label{sec:disc-granularity}

The evidence on the quality of the resulting tests is less conclusive than on the kind of work. The \ac{PR}-level smell estimates of RQ$_4$ are borderline, opposite in sign for additions and modifications, and not reproduced for individual test methods, so we treat them as hypotheses. If confirmed, they would imply that agent-supported \acp{PR} more often perform the activity in which their Smell Delta is lower, so that the net consequence for a project would depend on its mix of testing work. The divergence between the two levels admits non-exclusive explanations: smells defined on the test class or its shared setup, such as General Fixture, are counted at the \ac{PR} level but only partly attributed to test methods; a few contributions with very large deltas may dominate the \ac{PR}-level estimates; and the two levels weight \acp{PR} differently. Decomposing the deltas into introduced and removed occurrences, separating class-level smells, and assessing the influence of extreme contributions would distinguish them.

Two other results indicate where quality differences may lie, if any. Tests that lose their oracle are rare but concentrated in a few projects, in some of which agent-supported \acp{PR} lose oracles far more often (RQ$_3$), and in test-only \acp{PR} the tests that agents contribute cover less of the classes they exercise (RQ$_5$). Both rest on few repositories, and neither smells nor coverage measure whether tests would detect faults~\citep{inozemtseva2014}. More generally, estimates of the quality of agent-written tests can depend on the unit of analysis, so studies of this question should state the level at which their claims hold and, where possible, report more than one.

\subsection{Implications}
\label{sec:disc-implications}

\paragraph{For researchers.} Comparisons between agent-supported and human-authored contributions should be made within projects, or model project identity explicitly, since pooled comparisons conflate agent behavior with agent adoption (Section~\ref{sec:disc-within-repo}). The unit of analysis should be stated and justified, as contribution-level and method-level analyses of the same data can yield different estimates, and coverage studies should state which tests and which lines a measure attributes to a contribution, since designs that seem similar estimate different quantities (Section~\ref{sec:disc-accretive}). Finally, most current evidence concerns the tests that agents write, whereas our results suggest that the tests they do not update or delete are equally relevant to the evolution of test suites.

\paragraph{For practitioners.} In our data, the tests that agents add resemble those of developers in the same project, and we detect no difference in how well they cover the production changes they accompany. The main difference concerns the existing tests: teams adopting coding agents should not assume that their maintenance will follow from agent-supported work at the same rate as from human work, and may want to request it explicitly, for instance by asking agents to review and update the tests linked to the code they change. When reviewing agent-supported \acp{PR}, two checks are inexpensive: whether a modified test still contains an assertion, and, in \acp{PR} that only add tests, which part of the class under test the new tests exercise. Whether agent-supported changes to existing tests also deserve stricter review for test smells depends on the robustness of the RQ$_4$ estimates (Section~\ref{sec:disc-quality}).

\paragraph{For agent and tool builders.} The lower likelihood that agent-supported \acp{PR} modify or delete tests suggests a concrete design direction. Agent instructions and scaffolds could make test maintenance part of the task, for example by retrieving the existing tests linked to the production methods being changed and asking whether they should be updated or removed before new ones are written. Agents could also report which existing tests exercise the changed code and whether they were updated, which would make the omission visible to reviewers. Since smell detection and per-test coverage are inexpensive to compute on a \ac{PR}, both could be integrated into such reports, should the RQ$_4$ and RQ$_5$ differences prove robust.

\subsection{Threats to Validity}
\label{sec:threats-to-validity}

\paragraph{Construct validity.} Whether a coding agent participated in a \ac{PR} is inferred from the traces it leaves: we combine \ac{PR}-level and commit-level signals following~\citet{robbes2026c}, which reduces the reliance on any single convention but cannot be exhaustive. Agent use that leaves none of these traces places agent-supported \acp{PR} in the human-authored group, which we therefore describe as primarily human-authored. The direction of the resulting bias is uncertain. The heuristic that misclassification biases estimates toward the null, and that results are therefore conservative, holds only when the error is nondifferential with respect to the outcome and independent of errors in other variables, and it has several exceptions even then~\citep{yland2022misconceptions}. Neither condition can be taken for granted here: whether agent use leaves a trace plausibly depends on the agent, on the period, on project conventions, and on the type of contribution, all of which may also be associated with testing behavior. Since the misclassification rate may also differ between repositories, it can affect the repository-specific differences, and therefore $\hat{\tau}$, as well as the average effect. We therefore do not assume that our estimates are conservative. Conversely, an agent trace does not indicate how much of a \ac{PR} the agent produced or how much the developers revised it, so agent support is a binary approximation of a graded construct, and \acp{PR} matching several agents are treated as agent-supported without distinguishing between agents.

Test code is identified with a heuristic validated in our previous study (Section~\ref{sec:change-measurement}). Test changes are linked to production changes through statically resolved calls, so the number of co-evolving \acp{PR} is a lower bound, and changed test units that \textsc{Hamster} does not detect are excluded (Section~\ref{sec:rq-2-experimental-procedure}). Smell detectors have false positives and negatives, and the aggregate Smell Delta weights all categories equally. Neither smells nor coverage measure whether tests would detect faults, which would require measures such as mutation analysis.

\paragraph{Internal validity.} Our study is observational, and the differences we report are associations rather than causal effects of agent support. Developers may delegate to agents well-scoped, easily testable, or greenfield tasks, and our models adjust for the repository and change size but not for the type or difficulty of the task. Since size may itself be affected by agent support, our coefficients estimate the difference between \acp{PR} of equal size rather than the total difference associated with agent support. Finally, we analyze only merged \acp{PR}, so our results describe what projects accept after review, with changes made during review credited to the \ac{PR}, not the contributions that agents initially produce.

\paragraph{Conclusion validity.} We control the family-wise error rate with the Holm-Bonferroni method within each family of tests. The two aggregate Smell Delta models are analyzed outside these families, and with Holm correction across the four aggregate tests their \textit{p}-values (0.024 and 0.046) would not be significant, so we regard them as indicative. These models use an identity link on heavy-tailed signed differences of counts, so their estimates may be sensitive to extreme contributions. Where residual diagnostics indicate a misspecified dispersion, we do not rely on the width of the intervals, and the counts of repositories whose specific interval excludes 1 are descriptive and not adjusted for multiplicity. Non-significant results are not evidence of equivalence.

\paragraph{External validity.} We restrict the study to Java, whose standardized build tools make coverage measurement feasible at scale, and our findings should not be assumed to transfer to other ecosystems; RQ$_5$ is further limited to the repositories that can be built with Maven or Gradle, which may differ systematically from the others. Requiring at least ten agent-supported and ten human-authored \acp{PR} retains 391 of the 22,932 repositories with agent activity, and selects projects with both substantial agent adoption and sustained human contribution. Codex, Copilot, and Claude Code account for almost all agent-supported \acp{PR} (Section~\ref{sec:data-characterization}), so our results mainly characterize these three, and they describe the period from January 2025 to April 2026 rather than a stable state of a rapidly evolving practice.

\section{Conclusion and Future Work}\label{sec7}

This paper compared testing in agent-supported and human-authored \acp{PR} along five dimensions: the frequency and extent of testing, the type and context of test changes, the structure of test code, test smells, and code coverage. On 114,134 merged Java \acp{PR} from 391 repositories that contain both groups, we made every comparison within repositories, controlling for the size of the contribution.

Our results support three conclusions. First, comparisons of agentic and human contributions depend on the project: pooled across repositories, the two groups appear almost indistinguishable in how often they change test code, whereas within repositories agent-supported \acp{PR} do so more often, and the size of the difference varies considerably between repositories. Second, agent participation is associated with a different mix of testing work: more additions and fewer modifications and deletions of existing tests, without a detected difference in most structural characteristics of the added tests or in the coverage of production changes. Third, whether agent participation affects test quality remains open: the differences in test smells are borderline and depend on the level of analysis, and the lower coverage of agents' tests in test-only \acp{PR} rests on few repositories. For practitioners, this means that the maintenance of existing tests should not be assumed to follow from agent-supported work, and may need to be requested explicitly.

Several directions follow. A longitudinal analysis of test additions, modifications, and deletions over the history of each project would show whether the per-contribution difference leads to faster suite growth or shifts maintenance toward developers. Modeling repository-level moderators would help explain the variation between repositories. For test smells, decomposing the deltas into introduced and removed occurrences and reconciling the two levels of analysis would establish whether the borderline differences are robust, and a per-agent comparison would require a dataset in which agents other than the three most frequent ones are sufficiently represented. Separating what agents initially produce from what review corrects would make it possible to attribute the observed differences to the agent, to the developer, or to the review process, and mutation analysis would show whether coverage differences translate into differences in fault detection. Finally, replicating the study in other ecosystems and over a longer window would show how far these findings extend as coding agents and the practices around them evolve.

To support verification and reuse, the dataset, the queries, and the analysis scripts used in this study are publicly available in our replication package~\citep{milanese2026r}.

\section*{Declarations}

\textbf{Funding:} This publication is part of the project PNRR-NGEU, which has received funding from the MUR – DM 118/2023.
\\\\\textbf{Ethical Approval:} Not applicable.
\\\\\textbf{Informed Consent:} Not applicable.
\\\\\textbf{Author Contributions:} Conceptualization: All authors; Methodology: R. Milanese, F. Salzano, and H. Muradyan; Formal analysis and investigation: R. Milanese and F. Salzano; Writing - original draft preparation: All authors; Writing - review and editing: All authors; Funding acquisition, Resources and Supervision: R. Pareschi, F. Fasano, and M. Fazzini.
\\\\\textbf{Data Availability Statement:} All data and scripts are available in the study's replication package, published at: \url{https://doi.org/10.5281/zenodo.21914644}.
\\\\\textbf{Conflict of Interest:} The authors declare no conflict of interest.
\\\\\textbf{Clinical Trial Number:} Not applicable.


\bibliographystyle{spbasic}      
\bibliography{bibliography}

%
%

\end{document}